\documentstyle[epsf,draft,jgrga]{agu2001} 
\renewcommand{\epsilon}{\varepsilon}
\renewcommand{\phi}{\varphi}
\newcommand{\hpm}{\hphantom{-}}
\newcommand{\hpq}{\hphantom{?}}

\newcommand{\vex} { \vspace{1.0ex} }

\newcommand{\vnx} { \vspace{-2.5ex} }
\newcommand{\hs} {\hspace{-0.133em}}
\newcommand{\hbb}{\hspace{-0.5em}}

\newcommand{\hsp}{\hspace{0.5em}\hpq}
\newcommand{\beq}{\begin{equation}}
\newcommand{\eeqn}{\nonumber\end{equation}}
\newcommand{\eeq}[1]{\label{#1}\end{equation}}
\newcommand{\flo}[2]{\mbox{$#1 \kern-0.06em \cdot \kern-0.08em 10^{#2}$}}

\newcommand{\Frac}[2]{\frac{\displaystyle\strut #1}{\displaystyle\strut #2} }

\newcommand{\var}{ \mathop{ \rm var }\nolimits }

\newcommand{\dint}{\int\hspace{-0.28em}\int}

\newcommand{\dss}{\displaystyle}

\newcommand{\Der }[2]{\frac{ \partial \displaystyle\strut #1 }
                           { \partial \displaystyle\strut #2 } }
\newcommand{\ntab}[2]{ \multicolumn{1}{#1}{#2} }
\newcommand{\nntab}[2]{ \multicolumn{2}{#1}{#2} }

\newcommand{\nnnntab}[2]{ \multicolumn{4}{#1}{#2} }
\newcommand{\Deg}[1]{#1^\circ}

\ifjdraft
    \newcommand{\BegCorr}[1]{\marginpar{$\Longleftarrow$ \bf #1}}
    \newcommand{\EndCorr}[1]{\marginpar{$\longleftarrow$ \small #1}}
  \else
    \newcommand{\BegCorr}[1]{}
    \newcommand{\EndCorr}[1]{}
\fi
\authorrunninghead{PETROV AND BOY}
\titlerunninghead{ATMOSPHERIC PRESSURE LOADING}
\journalid{}
\articleid{}{}
\paperid{2003JB002500}
\cpright{agu}{2003}
\received{March 18, 2003}
\revised{October 29, 2003} 
\accepted{November 19, 2003} 
\published{}

\begin{document}

\title{Study of the atmospheric pressure loading signal in VLBI observations}
\author{Leonid Petrov}
\affil{NVI, Inc./NASA Goddard Space Flight Center, Greenbelt, Maryland, USA}
\authoraddr{Code 926, NVI, Inc./NASA Goddard Space Flight Center, 
            Greenbelt, MD 20771 USA. \mbox{(e-mail: Leonid.Petrov@gsfc.nasa.gov)} }

\author{Jean-Paul Boy}
\affil{Code 926, NASA Goddard Space Flight Center, Greenbelt, Maryland, USA}
\authoraddr{Code 926, NASA Goddard Space Flight Center, 
            Greenbelt, MD 20771 USA. \mbox{(e-mail: boy@bowie.gsfc.nasa.gov)} 
            }

\ifjdraft
 \par\bigskip\bigskip\bigskip\bigskip\par
 \mbox{ \LARGE\sf Accepted on November 19, 2003 }
 \par\bigskip\par
 \mbox{ \Large\bf  Manuscript \thepaperidnumber  }
\fi

\setcounter{secnumdepth}{2}

\begin{abstract}

  Redistribution of air masses due to atmospheric circulation causes loading 
deformation of the Earth's crust which can be as large as 20 mm for the 
vertical component and 3 mm for horizontal components. Rigorous computation 
of site displacements caused by pressure loading requires knowledge of the 
surface pressure field over the entire Earth surface. A procedure for 
computing 3-D displacements of geodetic sites of interest using a 6-hourly 
pressure field from the NCEP numerical weather models and the Ponte and 
Ray [2002] model of atmospheric tides is presented. We investigated possible 
error sources and found that the errors of our pressure loading time series 
are below the 15\% level. We validated our model by estimating the admittance 
factors of the pressure loading time series using a dataset of 3.5 million 
VLBI observations from 1980 to 2002. The admittance factors averaged over 
all sites are $0.95 \pm 0.02$ for the vertical displacement and 
$1.00 \pm 0.07$ for the horizontal displacements. For the first time 
horizontal displacements caused by atmospheric pressure loading have been 
detected. The closeness of these admittance factors to unity allows us to 
conclude that on average our model quantitatively agrees with the observations
within the error budget of the model. At the same time we found that the model 
is not accurate for several stations which are near a coast or in mountain 
regions. We conclude that our model is suitable for routine data reduction 
of space geodesy observations.

\end{abstract}

\begin{article}

\section{Introduction} 

   At the level of precision of modern space geodetic techniques the 
Earth's crust is not static, but deformable. The Earth's crust deformation can 
be caused by processes inside the Earth, by gravitational forces of external 
celestial bodies, by changes of the centrifugal potential, and by various mass 
loads. Analysis of geodetic observations made from the deformable surface 
of our planet requires applying a model of these deformations. The precision 
of this model should be comparable with the precision of the measurements, 
otherwise unaccounted site position variations due to crust deformation become
a factor which limits the accuracy of measurements, and the potential 
of geodetic techniques cannot fully be exploited.

   In this paper we focus on the Earth's crust deformation caused by the 
load of the atmosphere. As was found by E. Torricelli in 1644 \citep{r:torr},
the atmospheric pressure is not constant, but has variations at the level of
20--50 mbar. \citet{r:darwin} was the first who realized that this can cause 
deformation at the level of several centimeters, and he proposed a simple 
model for its computation. However, before the advent of space geodesy,
quasi-random site displacements at the level of centimeters were not 
directly measurable and, therefore, there was no necessity to take them 
into account.

   Rapid development of space geodetic techniques in the 1980-s made it 
feasible to try to detect atmospheric pressure loading signal from the 
measurements of site positions. \citet{r:truba} and later \citet{r:rab}, 
\citet{r:rabshu} and \citet{r:dampre} made quantitative assessments of the 
impact of passing cyclones and anticyclones on measurements of site positions 
assuming that the pressure distribution in cyclones or anticyclones can be 
described by a simple mathematical model. \citet{r:manabe} tried to find 
a correlation between predicted atmospheric pressure loading and the time 
series of site position determined from very long baseline interferometry 
(VLBI) observations during 1984--1989 at several stations, but had 
to acknowledge that ``the observationally determined vertical displacements 
is not explainable as being caused by the atmospheric loading, since the 
dispersion of the observed vertical displacements is too large''. Three years 
later \citet{r:damvlbi} and independently \citet{r:danatmload} succeeded 
in detecting atmospheric pressure loading signal using more sophisticated 
approaches. \citet{r:danatmload} estimated coefficients of linear regression 
between vertical site displacements and local pressure using VLBI data. They 
found that these coefficients for the majority of sites are in reasonable 
agreement with the coefficients derived by \citet{r:manabe}, and that applying 
an empirical model based on the regression between vertical site position and 
local pressure improves baseline length repeatability. \citet{r:damvlbi}, 
hereafter referred as VDH, used another approach. They analyzed the reduction 
of variance of the estimates of baseline lengths derived from analysis of the 
same VLBI dataset. They found that the reduction of variance is consistent 
with the hypothesis that only approximately 60\% of the computed pressure 
loading contribution is present in the VLBI length determination. Applying 
the same technique to global positioning system (GPS) data allowed 
\citet{r:gpsatmloa} to conclude that 57\% of the pressure loading signal is 
present in the baseline length residuals.

   Although the presence of atmospheric pressure loading signal was confirmed
in observations, modeling this signal did not come into practice for routine 
data reduction. First, a rigorous computation of displacements caused by mass 
loading requires handling a gigantic volume of information and enormous
processor power. Second, there was no certainty whether the model is correct.
Results of VDH and \citet{r:gpsatmloa} mean that observations did not confirm 
{\em quantitatively} the atmosphere pressure loading model. Without
solving this discrepancy, applying the atmosphere pressure loading model 
for processing routine observations is not warranted.

   There are several factors that motivated us to re-visit this topic. 
First, the National Centers for Environmental Prediction (NCEP) Reanalysis 
project \citep{r:ncep} now provides a continuous, uniform dataset of surface 
pressure on a $ \Deg{2}\!.5 \times \Deg{2}\!.5 $ grid with a 6 hour resolution 
for more than 40 years which was not available a decade ago. Second, rapid 
development of high speed networks and processor power makes it possible to 
retrieve and process voluminous meteorological data assimilation models in 
almost real time. Third, the accuracy of geodetic observations has increased 
considerably during the last ten years, which has improved our ability 
to detect subtle Earth's crust motions.

  The objective of our study was to develop a procedure of computing 
displacements caused by atmospheric pressure loading which is suitable for 
routine analysis of geodetic observations, and to compare these time series 
of pressure loading with a dataset of all VLBI observations from 1980 to 2002.
The purpose of this comparison is to get a quantitative measure of the 
agreement between the model and observations, and infer whether the model 
is correct or wrong. In order to do it, we thoroughly examine the error budget 
and on the basis of these estimates compute the expectation of the deviation 
of the observations from the model. Our goal is to determine whether 
the observations deviate from the expectation or not. If the agreement test 
deviates from the expectation at a statistically significant level, this 
means that either there is an error in computations or there is a fundamental 
flaw in our understanding of the physics of the phenomena under consideration. 
Then the  model must be rejected at this point. If the outcome of the 
statistical test is within the predicted range based on known deficiencies 
of the model, this means that the procedure for computation of the atmosphere 
pressure loading can be accepted for routine reduction of observations.

%
%

  In the second section of the paper we describe our method of computing site 
displacements caused by pressure loading and assess the error budget of our 
calculations. We re-analyzed a dataset of 3.5 million VLBI observations 
and performed several statistical tests of agreement between the model of 
pressure loading and the observations. The method of data analysis is described
in section~3. The results of the VLBI data analysis are discussed in section~4.
Concluding remarks are given in section~5. An efficient procedure for 
computing a time series of pressure loading is outlined in the Appendix.

\section{Computation of Displacements Using Meteorological Models} 

  According to \citet{r:farrell} the vertical displacement at a station 
of coordinates $\vec{r}$ induced by surface pressure variations 
$\Delta P(\vec{r}\,',t)$ is equal to:
\beq 
  u_r(\vec{r},t) =  \dint \Delta P(\vec{r}\,',t) \, G_{\sc R}(\psi) \cos \phi' 
                    d \lambda' d \phi'
\eeq{e:b1}
The vertical Green's function is 
\beq
   G_{\sc R}(\psi) = \frac{f a}{g_0^2} \sum_{n=0}^{+\infty} h^{'}_n P_n 
                     (\cos \psi)
\eeq{e:b2}
 $f$, $a$ and $g_0$ are the universal constant of gravitation, the mean Earth's 
radius and the mean surface gravity as defined in PREM \citep{r:PREM}, 
$\phi'$ is the geocentric latitude, and $\lambda'$ is the longitude. 
$\psi$ is the angular distance between the station with coordinates 
$\vec{r}$ and the pressure source with coordinates $\vec{r}\,'$. $P_n$ is 
the Legendre polynomial of degree~$n$.

The horizontal displacement is computed this way:
\beq
  \vec{u}_h(\vec{r},t) = \dint \vec{q}(\vec{r},\vec{r}\,') \, 
            \Delta P(\vec{r}\,',t) \, G_{\sc h}(\psi) 
            \cos \phi' d \lambda' d \phi'
\eeq{e:b3}
  where $\vec{q}(\vec{r},\vec{r}\,')$ is the unit vector originating from the 
station, tangential to the Earth's surface, which lies in the plane determined 
by the radius-vectors to the station and to the pressure source. 
The tangential Green's function is \citep{r:farrell}: 
\beq
    G_{\sc h}(\psi) = -\frac{f a}{g_0^2} \sum_{n=1}^{+\infty} l^{'}_n 
                \frac{\partial P_n (\cos \psi)}{\partial \psi}
\eeq{e:b4}
  Numerical evaluation of the Green's function requires the computation of load
Love numbers $h^{'}_n$ and $l^{'}_n$ up to a high spherical harmonic degree 
($n=9000$ in this study) for a spherically symmetric, non-rotating, elastic 
and isotropic (SNREI) Earth model. The method of numerical computation 
of Green's functions is presented by \citet{r:farrell}.  
 
%
%
%

  We model the oceanic response to atmospheric pressure forcing as an inverted 
barometer (IB):
\beq
      \Delta P_a + \Delta P_w - \Delta \bar{P}_o = 0
\eeq{e:b5}
  where $\Delta P_a$ is the variation of local atmosphere pressure, 
$ \Delta P_w $ is the local variation of the ocean bottom pressure due 
to induced sea level change, and $ \Delta \bar{P}_o $ is the mean atmosphere 
pressure over the world's oceans:
\beq
   \Delta \bar{P}_o = 
          \Frac{\dint\limits_{\hbb\hbb\: ocean} \Delta P(\vec{r}\,',t) 
               \cos \phi' d \lambda' d \phi'}
               {\dint\limits_{\hbb\hbb\: ocean} \cos \phi' d \lambda' d \phi'}
\eeq{e:mass}
which is applied uniformly at the sea floor \citep{r:dampre}. This term is 
introduced in equation~\ref{e:b5} in order to enforce conservation of ocean 
mass. Thus, the total ocean bottom pressure, $ \Delta P_a + \Delta P_w $, 
is described by equation~\ref{e:mass}. It has been shown in numerous studies 
(see, for example, \citep{r:tierney}) that this model adequately describes 
the sea height variations for periods longer than 5--20 days. However, the 
ocean response significantly deviates from the IB hypothesis for shorter 
periods \citep{r:wunsch}. 

   Since $\Delta \bar{P}_o$ is zero over the land and depends only on time 
over the world's oceans, it is convenient to split integrals \ref{e:b1} and 
\ref{e:b3} into a sum of integrals over the ocean and over the continental 
surface. In our computation we use the land-sea mask from the 
FES99 \citep{r:mask} ocean tidal model with a $ \Deg{0}\!.25 $ spatial 
resolution. A practical algorithm for the computation of site displacements 
caused by atmospheric pressure loading is presented in the Appendix.

   Since the NCEP Reanalysis numerical weather models have a time resolution 
of 6 hours, the semidiurnal ($S_2$) atmospheric tide induced by solar heating 
cannot be modeled correctly, because its frequency corresponds exactly to the 
Nyquist frequency. The diurnal ($S_1$) atmospheric tide is somewhat distorted 
as well, because of the presence of the ter-diurnal signal, which is folded 
into the diurnal frequency due to sampling. This problem was investigated 
by \citet{r:dool} in detail, who proposed a temporal interpolation procedure 
which to some extent allows one to overcome the problem. Based on this approach 
\citet{r:ponteray} recently developed a gridded global model of the $S_1$ and 
$S_2$ atmospheric tides with a spatial resolution of
$ \Deg{1}\!.125 \times \Deg{1}\!.125 $. For reasons discussed by these authors, 
we believe their model better represents the atmospheric tides than the diurnal 
and semidiurnal signal which is present in the NCEP Reanalysis model. Using 
these maps we have computed amplitudes and phases of the loading caused by 
diurnal and semidiurnal atmospheric tides for VLBI and satellite laser 
ranging (SLR) sites. 

   For each grid point we have estimated four parameters using least squares
(LSQ): mean pressure, sine and cosine amplitude of the $S_1$ signal, and cosine
amplitude of the $S_2$ signal in the surface pressure field over the time 
period from 1980 to 2002. This four-parameter model is subtracted at each 
point of the grid from the NCEP Reanalysis pressure field before evaluation 
of the convolution integral. Thus, our time series has zero mean and no signal 
at $S_1$ and $S_2$ frequencies. The total loading is the sum of the time series 
and the harmonic model of the $S_1$ and $S_2$ loading caused by atmospheric 
tides.
  
\subsection{Characteristics of the Atmospheric Pressure Loading Displacements }

   In \callout{figures~\ref{f:wettzell} and \ref{f:hartrao}} 
we show examples of time series of the modeled displacements for the 
period 2000--2003 and their power spectrum at the Wettzell and Hartrao 
stations respectively. These stations are representative of mid-latitude and 
equatorial inland stations. 

   All station displacements show significant narrow-band diurnal and 
semidiurnal signals. The displacements for low-latitude stations are 
characterized by a strong wide-band annual and semi-annual signals and 
relatively weak signal for periods below 10 days, except strong peaks at the
$S_1$ and $S_2$ frequencies. Mid-latitude stations show just the opposite 
behavior. For the mid-latitude regions circulation of low and high pressure 
structures with periods of 5--10 days is typical. These periods are also 
at the edge of the validity of the IB model for describing the oceanic 
response to atmospheric pressure forcing.

\begin{figure}
   \caption{a) Vertical and b) north displacement induced by atmospheric 
            pressure loading at the station Wettzell. Power spectrum of the
            c) vertical and d) north displacement. }
   \label{f:wettzell}
\end{figure}

\begin{figure}
   \caption{a) Vertical and b) east displacement induced by atmospheric 
            pressure loading at the station Hartrao. Power spectrum of the
            c) vertical and d) east displacement. }
   \label{f:hartrao}
\end{figure}

\begin{figure}
   \caption{Temporal autocorrelation of the vertical component of 
            the atmospheric pressure loading at the station Algopark.}
   \ifjdraft \else 
       \hspace{-7mm}
       \mbox{ \epsfxsize=83mm \epsfclipon \epsffile{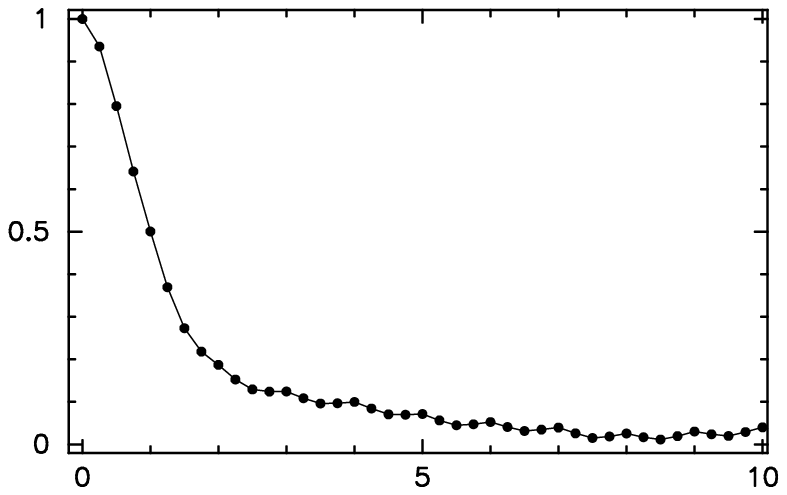} }
   \fi
   \label{f:tcorr}
\end{figure}

\begin{figure}
   \caption{Smoothed spatial autocorrelation of the vertical component 
            of the atmospheric pressure loading.}
   \ifjdraft \else 
       \hspace{-7mm}
       \mbox{ \epsfxsize=83mm \epsfclipon \epsffile{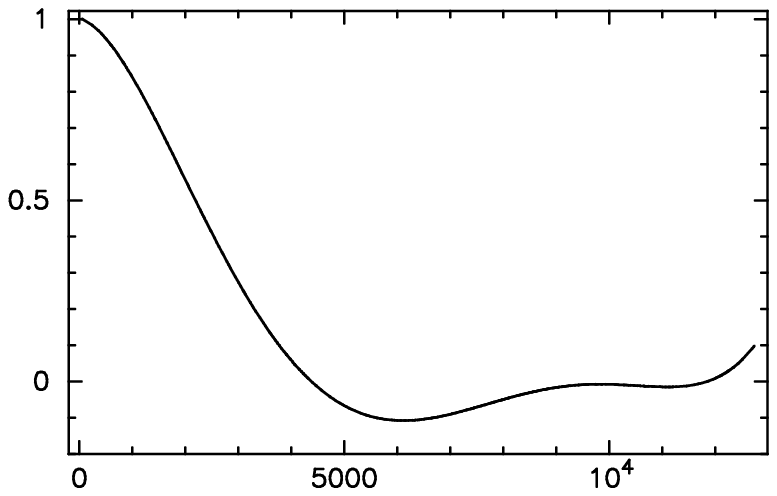} }
   \fi
   \label{f:scorr}
\end{figure}

  The rms of vertical and horizontal displacements is presented in columns 4 
and 5 of \callout{table~\ref{t:statist}}. \callout{Figure~\ref{f:tcorr}} 
shows an example of the temporal autocorrelation function for the vertical 
displacement for the mid-latitude station Algopark. The autocorrelation 
rapidly drops to the level less than 0.2 for time intervals longer than 2 days.
The smoothed spatial autocorrelation of the vertical displacement induced 
by atmospheric pressure loading is presented in \callout{figure~\ref{f:scorr}}.
We see that the correlation for baselines shorter than 1000 km is very 
high, typically greater than 0.9, and only for baselines longer than 
3000 km does it drop below the level of 0.2 .

\subsection{Error Budget \label{s:err}}
   
  There are four major sources of errors in the computation of site 
displacements caused by atmospheric pressure loading: 1) errors in the 
Green's functions, 2) errors in the land-sea mask, 3) errors in the pressure 
field, and 4) mismodeling the ocean response to atmospheric pressure forcing.

  The Green's functions are computed for a SNREI Earth model adopting PREM 
elastic parameters. Thus, we neglect the effects induced by Earth's 
anelasticity and ellipticity. The differences between our Green's functions 
and Green's functions for an anelastic Earth model (see for 
example \cite{r:pagiatakis} or \cite{r:okubo}) are typically below 1--2\%. 
The effect of the Earth's ellipticity is of the order of magnitude of the 
Earth's flattening, i.e. 0.3\%.

\begin{figure}
   \ifjdraft \else 
        \hspace{-6mm}
        \mbox{ \epsfxsize=83mm \epsfclipon \epsffile{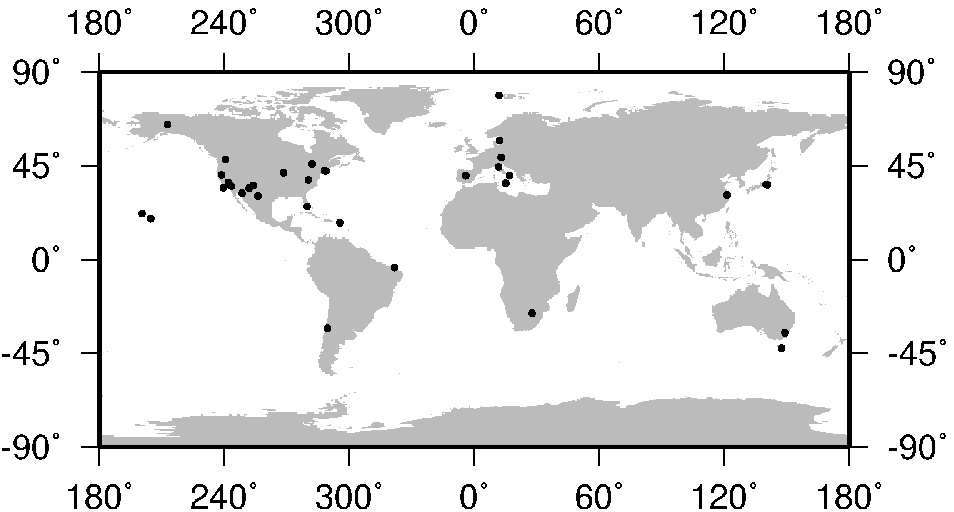} }
   \fi
   \caption{Map of VLBI stations used in the analysis.}
   \label{f:map}
\end{figure}

  Since the $ \Deg{2}\!.5 $ spatial resolution of the NCEP Reanalysis surface 
pressure field is not sufficient to correctly represent the coastline, we 
chose a land-sea mask with a $ \Deg{0}\!.25 $ resolution from the
FES99 \citep{r:mask} ocean tidal model. The land-sea mask and the station 
distribution are shown in \callout{figure~\ref{f:map}}. We assume that 
enclosed and small semi-enclosed seas (such as Baltic, Black, Red, Caspian
Seas and Persian Gulf) respond to atmospheric pressure forcing as a 
"non-inverted barometer", i.e. atmospheric pressure variations are fully 
transmitted to the sea floor. In order to evaluate the effect of the errors in 
the land-sea mask, we computed the time series of atmospheric pressure loading 
with both the $ \Deg{0}\!.25 $ and $ \Deg{0}\!.50 $ land-sea masks. 
The differences between these two estimates are typically  about 5\%. 
%
%
It is worth mentioning that the difference between  the loading estimates with 
the $ \Deg{2}\!.5 $ and $ \Deg{0}\!.25 $ land-sea masks can reach 10\% for the 
vertical component and  30\% for the horizontal components, even for a station 
like Wettzell (Germany) which is 500 km from the coast.

  Another source of error in our computations are errors in the surface 
pressure field from the NCEP Reanalysis. One way to compute this is to compare
directly the difference between the model and surface pressure observations. 
\citet{r:velicogna} presented estimates of the rms differences for two 
different regions (Arabic Peninsula and United States). On the basis of their 
estimates we conclude that the errors of the NCEP surface pressure field 
on these selected areas are at the level of 5\%. 

  Another measure of possible errors in the surface pressure field model
is the difference between two variants of NCEP numerical weather models: 
the NCEP Reanalysis and the NCEP Operational Final Analysis, 
although these two models are obviously not independent. The NCEP Operational
Final Analysis models is an improved version with respect to an earlier model 
NCEP Reanalysis. The improved horizontal ($\Deg{1}$ instead of $\Deg{2}\!.5 $) 
and vertical (42 layers instead of 28) resolution allows a better modeling of
atmospheric dynamics. We computed station displacements due to atmospheric 
pressure loading using the NCEP Operational dataset with a spatial resolution 
of $ \Deg{1}\!.0 $ for the period from April 2002 to January 2003. The rms of 
the differences between the vertical and horizontal displacements computed 
with the NCEP Reanalysis and the NCEP Operational data are shown in the 6th 
and 7th columns of table~\ref{t:statist}. Correlations between the differences 
and the modeled signal are shown in columns 8, 9. The mean error is about 10\% 
for the vertical and horizontal components. It is noticeable that the 
differences are larger for stations enclosed by mountains, for example, 
Santiago. This is due to the fact that the $ \Deg{2}\!.5 $ or even 
$ \Deg{1}\!.0 $ spatial resolution of the NCEP Reanalysis and Operational 
datasets is not sufficient to model the topographic variations in mountainous 
areas and, therefore, the surface pressure variations. Our results are similar 
to conclusions made by \citet{r:velicogna}. The precision of computation of 
pressure loading displacements is worse for mountainous stations.

  In order to evaluate the errors caused by mismodeling of the ocean response, 
we compared ocean bottom pressure variations, as well as the induced loading 
effects, from two runs of the Coupled Large-scale Ice Ocean (CLIO) general 
circulation model \citep{r:clio}; the first one is forced by atmospheric 
pressure, surface winds and heat fluxes \citep{r:deviron}, and the other one 
is forced only by surface winds and heat fluxes, i.e. assuming an IB response. 
The differences between these two runs can therefore be interpreted as the 
departure of the ocean response from the IB assumption. We also validated the
bottom pressure changes modeled by CLIO with measurements from the Global 
Undersea Pressure (GLOUP) data set 
\mbox{\tt [http://www.pol.ac.uk/psmslh/gloup/gloup.html]} and found that the
CLIO model agrees with the measurements of the bottom pressure at the level of 
20\%. We show in the 10th and 11th columns of table~\ref{t:statist} the ratio 
of the rms of these differences to the rms of our atmospheric loading series 
with the IB model for all stations, as well as a mean value of this ratio. 
Correlations between the differences and modeled signal are shown in columns 
12 and 13. Thus, the mean vertical and horizontal errors are below 10\% and 
20\% respectively. As expected, these values are higher for island stations 
(f.e., Kokee, Mk-vlba) or stations close to the coasts (Richmond, Hobart26, 
etc.), where the atmospheric loading itself is very small: rms below 1 mm and 
0.5 mm for the vertical and horizontal components respectively.

  \callout{Table~\ref{t:errbud}} summarizes the error budget. Combining all 
known sources of errors we evaluate the total uncertainty of our computation of 
site displacements due to atmospheric pressure loading to be 15\%.

\section{Validation of the Model Using VLBI Observations}

   We selected VLBI for validation of our time series of atmospheric
pressure loading. Each of the three main space geodetic techniques, GPS, 
SLR and VLBI, has its own advantages and disadvantages, although in
general they are quite competitive. We chose VLBI because of the maturity 
of the VLBI data analysis technique. Complete re-analysis of the whole set 
of VLBI observations takes about a couple of hours on rather a modest computer. 
Therefore, the consistency of reduction models and parameter estimation can 
easily be enforced. 
%
%
These factors make VLBI attractive for investigating 
tiny effects like atmospheric pressure loading.

\subsection{Observations}

  All dual-band Mark-3/Mark-4/K-4 VLBI observations carried out under various 
geodetic and astrometric programs from 1979 to the present are available 
on-line at the International VLBI Service for Geodesy and Astrometry (IVS) Data
Center at \mbox{\tt http://ivscc.gsfc.nasa.gov} \citep{r:ivs}. The VLBI data 
set has substantial spatial and time inhomogeneity. Typically, observations 
are made in sessions with a duration of about 24 hours. Observations were 
sporadic in the early 80s, but in January 1984 a regular VLBI campaign for the 
determination of EOP started first with 5-day intervals, from May 1993 with 
weekly intervals, and from 1997 twice per week. In addition to the 
observations dedicated to EOP measurements, various other observing 
campaigns were running. On average 150 sessions per year have been observed 
since 1984.

  In total 144 stations participated in observations, although a majority 
of them observed only during short campaigns. The stations which participated 
in more than $20\,000$ observations for more than three years were used for 
analysis. 46 stations satisfied these criteria. Four stations which 
participated in the Key Stone Project (KSP) \citep{r:ksp}, Kashim11, Koganei, 
Miura, Tateyama, were excluded since they observed mainly in a small local 
network, as well as two other stations, Crimea, because it had bad performance,
and Ylow7296, since its sensitivity was too low. Only observations on the 
baselines between the 40 strong selected stations were used, and other 
observations ($ \sim $6\%) were discarded. Sessions with less than 3 strong 
stations were discarded entirely. 3073 sessions from April 1980 to
December 2002 with more than 3.5 million observations remained, and they 
were used in the analysis. 

  The number of participating stations in each individual session varies from 
2 to 20, although 4--7 is a typical number. No station participated in all
sessions, but every station participated in sessions with many different 
networks. All networks have common nodes and, therefore, are tied together. 
Networks vary significantly, but more than 70\% of them have a size exceeding
the Earth's radius.

\subsection{Choice of Parameterization}

   The scatter of daily estimates of site positions is greater by a factor 
of 2--5 than the rms of atmospheric pressure loading displacements.
Therefore, we cannot directly see the signal in the site position time 
series. Besides, in order to resolve rank deficiency of the problem of 
simultaneous adjustment of coordinates of all site and EOP, net-rotation and 
net-translation constraints should be applied or an equivalent technique
should be used. As a result adjustments of site positions become linearly 
dependent. Since only a small number of stations participates in a typical 
VLBI experiment, the influence of position variations of other observing 
stations of the network on position variations of a station of interest 
is not diluted to a negligible level. It makes the interpretation of daily 
VLBI site position time series uncertain.

   One of the ways to assess validity of the model is to compute two time 
series of baseline lengths: the first with applying the atmospheric pressure 
loading model and the second without applying the model. Baseline lengths are 
invariant with respect to a rotation and translation and, therefore, 
net-translation and net-rotation constraints do not affect them. We introduce 
the reduction of variance coefficient $R$:
\beq
       R = \Frac{\Delta \sigma^2 + \sigma_m^2}{2 \, \sigma_m^2}
\eeq{e:e1}
  where $\Delta \sigma^2$ is the difference between the mean square of baseline 
length residuals before and after applying the model, and $ \sigma_m $ is the 
variance of the signal in the model. If the model is perfect and the signal 
under consideration {\em is not correlated} with another unmodeled effects, 
the coefficient is 1. If the baseline length series does not contain the 
signal coherent with the model at all, then applying the model increases 
the variance by the amount of the variance of the signal in that model, and 
the coefficient of reduction of variance is 0. We should emphasize importance 
of the assumption of the lack of correlation between the pressure loading 
signal and noise: we can extract the signal which is below the noise level 
if, and only if, some additional information about the noise is exploited.
The validity of this assumption is based on the fact that as can bee seen in 
figures~\ref{f:wettzell} and \ref{f:hartrao}, the spectrum of the atmosphere 
pressure loading at frequencies below one day is flat, except for a peak at 
the annual frequency for some stations. Thus, the atmospheric pressure 
loading can be considered to some degree as a stochastic Gaussian process. 
The spectrum of the VLBI residuals is also flat. Therefore, our condition
of lack of correlation between the atmosphere pressure loading series and 
residual unmodeled effects is fulfilled if an unmodeled contribution to VLBI 
delay and pressure loading are independent.

  Although this approach gives us a quantitative measure of the adequacy of 
the model, it has some disadvantages. It lets us determine only the 
reduction of variance coefficients for the projection of the difference of the 
site displacements vectors on the baseline vector instead of the coefficients 
for each site and each component independently. 

  Another approach is to represent the atmospheric pressure loading signal 
as a product $A \cdot a_m$, where $a_m$ is the modeled signal, and to estimate
directly from the VLBI time delays the unknown parameter $A$ which hereafter 
we call admittance factor between the modeled signal and the observables.
It can easily be verified that if $A$ is the only estimated parameter and 
the modeled signal is {\em not correlated} with an unmodeled contribution 
to the observable, then the expectation of the LSQ estimate of the admittance 
factor, $E(\hat{A})$, is
\beq
     E(\hat{A}) = \rho \Frac{\sigma_l}{\sigma_m}
\eeq{e:e2}
where $\rho$ is the correlation coefficient between modeled and true 
atmospheric pressure loading, and $\sigma_l$ is the variance of the true 
pressure loading. Under the same assumption the reduction of variance 
\mbox{$\Delta \sigma^2 = \biggl( 2 E(\hat{A}) - 1 \biggr) \sigma_m^2 $} and 
therefore, the coefficient of reduction of variance $R$ is equal to 
$E(\hat{A})$. If other parameters are adjusted in addition to $A$, this 
property is preserved at the level of correlations between $\hat{A}$ and the
estimates of other parameters. The admittance factor $A$ shows how much
of the power of the modeled signal is present in the observables.

\subsection{Estimation Model}

   We made solutions of two classes: global solutions, in which we estimated 
the positions and velocities of all sites over the entire data set, and 
baseline solutions, in which we estimated site positions for each VLBI 
experiment independently using ionosphere free linear combination of group 
delays at S and X bands. Estimated parameters were split into two classes: 
basic parameters, which are usually adjusted in processing VLBI experiments, 
and specific parameters of interest. Basic parameters belong to one of the 
three groups: 
\begin{itemize}
       \item [---]{\bf global} (over the entire data set): positions of 
                          511 primary sources and proper motions of 79 sources.
       \item [---]{\bf local}  (over each session):  pole coordinates and their
                          rates, UT1, UT1 rate; positions of other sources, 
                          azimuthal troposphere path delay gradients for all 
                          stations and their rates, station-dependent clocks 
                          modeled by second order polynomials, 
                          baseline-dependent clock offsets.
       \item [---]{\bf segmented} (over 0.33--1.0 hours): coefficients of 
                          a linear spline which models atmospheric path delay
                          (0.33 hour segment) and clock function 
                          (1 hour segment) for each station. The estimates 
                          of clock function absorb uncalibrated instrumental
                          delays in the data acquisition system.
\end{itemize}

  The rate of change of atmospheric path delay and clock function between two 
adjacent segments was constrained to zero with weights reciprocal to $40\, 
\mbox{psec/hour}$ and $\flo{2}{-14}\, \mbox{sec/sec}$ respectively in order 
to stabilize solutions. No-net-translation and no-net-rotation constraints 
were imposed on the adjustments of site positions and velocities as well
as no-net-rotation constraints were imposed on adjustments of source positions
in order to solve the LSQ problem of incomplete rank.

  A pair of baseline solutions was made with only basic parameters estimated
and station positions treated as local parameters. In reference solution B1 the 
atmospheric pressure loading time series was not applied; in solution B2 
the contribution due to the model of atmospheric pressure loading was added to
the theoretical model of the observables.

  In solutions G1, G5, G6 and G7 the list of global parameters included site 
positions, site velocities and overall admittance factors for the Up, East and
North components of the modeled atmospheric pressure loading signal for all 
sites combined. In solution G2 we estimated the set of the admittance factors 
for the Up, East and North components for each station independently, and other
parameters were the same as in the G1 solution.

  It should be noted that unlike estimation of {\em positions} of all
stations, estimation of the Up, East and North {\em admittance factors} for all 
stations of a global network does not result in a rank deficiency. A partial 
derivative of time delay $\tau$ with respect to the admittance factor of the
i'th station is given by $ \Der{\tau}{A_i} = \Der{\tau}{r_i} \, a_{mi}$,
where $r_i$ is a vector of station coordinates. As was shown in 
figure~\ref{f:tcorr}, the correlation between the atmosphere pressure loading 
at different stations is less than 0.1 at distances greater than 4000 km.

\subsection{Theoretical Model}

   The computation of theoretical time delays, with some exceptions, generally 
follows the procedure outlined in the IERS Conventions \citep{r:iers96} 
and described in more detail by \citet{r:masterfit}. The GOT00 model 
\citep{r:got99} of diurnal and semidiurnal ocean tides, the NAO99 model
\citep{r:nao99} of ocean zonal tides, the equilibrium model of the pole
tide and the tide with period of 18.6 years were used for the computation of
displacements due to ocean loading. The hydrology model of \citet{r:millya}
was used for the computation of displacements due to hydrology loading 
(D.S.~MacMillan and J.-P.~Boy, Effect of mass loading signals on crustal 
displacements measured by VLBI, manuscript in preparation; D.S~MacMillan, 
personal communication 2003). The empirical model of high frequency Earth 
orientation parameters derived from analysis of the VLBI data \citep{r:peteop} 
and the IERS96 semi-empirical nutation expansion were used. The \citet{r:niell} 
mapping functions were used for modeling and estimating the tropospheric path 
delay. The displacement of the reference point of a VLBI station due to thermal
expansion of the antenna was not modeled. Although \citet{r:termexp} proposed 
a model for an antenna's thermal expansion, attempts to validate this model 
were not successful. 
\BegCorr{C1}
We did not include the model of non-tideal ocean loading, because this effect
is one order of magnitude smaller than the atmosphere pressure loading, and
it has not been fully investigated. It will be considered in a future paper.
\EndCorr{C1}

\section{Discussion}

\subsection{Analysis of Global Admittance Factors}

  \callout{Table~\ref{t:gloadm}} shows admittance factors determined in 
solution G1. The uncertainties of the results were derived by propagating the 
group delay errors. These errors were computed on the basis of the signal 
to noise ratio of the cross correlation function of the recorded signal from 
the receivers and the empirical baseline-dependent reweighting parameters 
which, being added in quadrature to the uncertainties of group delays, made 
the $\chi^2\hs$/f of the postfit residuals close to unity. In addition to 
that, the formal uncertainties were scaled by a factor of 1.5. Numerous tests 
of splitting the dataset into subsets in time and in space showed that the 
VLBI formal uncertainties are underestimated by the factor of 1.5. The origin 
of this discrepancy is not completely understood, but there are indications 
that it may be caused by fringe phase variations due to instrumental 
errors \citep{r:petinst}.

   Above we showed the estimates of the rms of possible errors in our 
computation of atmospheric pressure loading and their correlations with the
modeled signal. Assuming that the different error sources a) are not 
correlated with each other, b) are small with respect to the signal, we can 
present the expectation of the admittance factor in this form:
\BegCorr{C6} 
\beq
    E(A) = 1 \: - \: \dss\sum k_i \, r_i \: + \: 
           \sum k_i^2 \, ( 2 r_i^2 - 1) \: + \: 
           4 \sum \sum_{i \neq j} k_i k_j r_i r_j \: + \: 
           O(k^3)
\eeq{e:e3}
\EndCorr{C6} 
   where $k$ is the ratio of the rms of errors to the rms of the modeled
signal, $r$ is the correlation between the error and modeled signal. The 
summing is done over all considered sources of errors.

   Using numerical values for $r_i$ and $k_i$ listed in table~\ref{t:errbud}, 
we evaluate the expectation of $A$: 0.90 . Our estimates of the admittance 
factors are close to this value. This means that the known deficiency of 
the model is sufficient to explain the small deviation of the estimates of 
the global admittance factors from~1.

   Since the typical spectrum of atmospheric pressure loading shows peaks at 
semi-diurnal (\flo{1.46}{-4} rad/s), diurnal (\flo{7.29}{-5}), semi-annual 
(\flo{3.98}{-7} rad/s) and annual frequencies (\flo{1.99}{-7} rad/s) 
(figures~\ref{f:wettzell}--\ref{f:hartrao}), we would like to see how 
applying the atmospheric pressure loading model affects residual harmonic site 
position variations at these frequencies. In order to assess this effect, we
made two solutions, G3 and G4, and estimated the sine and cosine amplitudes 
of position variations of all sites at 32 tidal frequencies, including 
these four frequencies with noticeable atmospheric pressure loading signal. 
Atmospheric pressure loading was applied in solution G4, but was not 
in solution G3. Amplitudes at the frequencies of diurnal, semidiurnal and long 
period bands where no tidal signal is expected were estimated as well in 
order to calibrate the uncertainties of the results. \BegCorr{C2} The ratio 
of the weighted sum of squares of the residual amplitudes over all stations 
at the specific frequency to its mathematical expectation, $\cal{P}$, was 
used as a measure of the power of the residual signal. In the absence of 
the signal these statistics should be less than 1.25 at the 95\% confidence 
level. Therefore, large values of $\cal{P}$ which exceed this limit indicate 
the presence of the residual signal. This technique is explained in more 
details by \citet{r:harpos}.

   \callout{Table~\ref{t:chi}} shows the estimates of $\cal{P}$ for each 
of the four frequencies of interest. \EndCorr{C2} We see that applying the 
model of atmospheric pressure loading reduces the amplitude of the residual 
signal at semi-diurnal and semi-annual frequencies, but noticeably increases
the amplitude at the annual frequency. In all cases the power of remaining 
residual signal is still significant. This table shows us that the atmosphere 
pressure loading is not the dominating source of observed residual harmonic 
site position variations at these frequencies.

   The presence of the narrow-band residual signal to some degree violates the
assumption of lack of correlations which was put at the basis of the agreement 
test. Two harmonic signals may be independent and correlated. In order to 
investigate the effect of the atmospheric pressure loading signal at the 
annual frequency, the frequency with the largest harmonic signal, we passed the 
pressure loading time series through a narrow-band filter with a passing 
window around the annual frequency: $[\flo{1.79}{-7}, \flo{2.19}{-7}]$ rad/s. 
We made two solutions, G5 in which we estimated the global admittance factor 
of the annual constituent of the atmospheric pressure loading and solution G6 
in which we estimated the global admittance factor with the signal at the 
annual band removed. The results are presented in table \ref{t:gloadm}. 
The admittance factor in solution G5 increased by 3\%. It gives us a measure 
of the distortion of the statistical tests caused by unaccounted correlations 
between the annual narrow-band unmodeled signal and atmospheric pressure 
loading. Similar results were reported by VDH who noted that removal of the 
annual signal from the atmospheric pressure loading increased variance 
reduction of the baseline length series.

   The fact that modeling atmospheric pressure loading increases the level 
of the residual signal at annual frequency does not necessarily mean that 
the model is wrong. If two anticorrelated signals contribute to the 
observables, then including the model of only one of the signals in the 
procedure of reduction of observations may result in increasing variance, even 
if the model is perfect. We know that various phenomena may contribute to the 
annual site position variation, such as hydrological signal, non-tidal ocean 
loading, thermal antenna deformation, mismodeling troposphere path delay, etc, 
and atmospheric pressure loading is not the greatest contributor. It should be 
noted that this means that a reduction of the variance coefficient and an 
admittance factor cannot be considered as a valid test of goodness of the 
model in this specific case. 

   We should acknowledge that currently we are unable to test {\em directly} 
whether the annual constituent of the atmospheric pressure loading signal is 
modeled correctly or not. It will be possible in the future when a complete 
model of site position variations will be built and the power of the residual 
signal will become less than the power of annual pressure loading signal. 
Although attempts to model seasonal effects show certain improvement 
\citep{r:dong}, we are still far from a solution of this problem.

   At the same time the admittance factor for the vertical displacements is 
close to unity at the level of measurement noise for the wide-band component 
of the modeled displacements due to atmosphere pressure loading after 
subtraction of the annual component. It provides us {\em indirect} evidence 
that we have modeled atmosphere pressure loading correctly at the annual 
frequency as well, since the Green's function and land-sea mask are frequency 
independent, and our estimate of the error budget set the upper limit of 
possible seasonal errors of the atmosphere pressure field.

\subsection{Analysis of Admittance Factors for Individual Stations.}

  Although we concluded in the previous section that the average admittance
factor is very close to unity, it does not necessarily mean that the 
admittance factors are close to unity for each individual station. 
\callout{Table~\ref{t:g2}} shows the estimates of the admittance factors from
the G2 solution. The estimates with the formal uncertainties greater 
than 0.5 are omitted in the table. It is not surprising that the admittance 
factor of the Kokee station is far from unity, even negative, since the 
station is located on an island in the middle of the Pacific ocean. It is more 
surprising that the admittance factors at the Hn-vlba and Westford stations 
are so different, since the distance between these stations is only 54 km and, 
indeed, the time series of the pressure loading are very similar. We suspect 
some problem with data at the station of Hn-vlba. Low admittance at the 
La-vlba and Pietown stations located in mountainous regions may be explained 
by greater errors in the model of surface pressure. We do not have 
explanation of the negative admittance factor at the station of Ov-vlba. 
Another peculiarity is the anomalously high admittance factors at the 
Seshan25 and Tsukub32 stations. It is known that the response of the Yellow 
Sea to tidal forcing 
is amplified by 10 times, and a significant non-linear $M_4$ tide is observed 
\citep{r:yellowsea}. We can expect that the response to atmospheric forcing 
will be also substantially non-linear and not consistent with the IB 
hypothesis. 

%
%

\subsection{Analysis of Reduction of Variance Coefficients}

  In order to compare our results with the early VDH paper, we performed our 
analysis in a manner similar to that used in the analysis of those authors. 
We computed the time series of baseline lengths in the B1 and B2 solutions. 
A linear model was fit in the series with discontinuities at epochs 
of seismic events for several stations. The weighted root mean square 
of residual baseline lengths was computed for all baselines with more than 
100 sessions for both the B1 and B2 solutions. 69 baselines satisfy this 
criterion. The coefficients of reduction of variance were computed using 
baseline length variances. 

  The histogram of the distribution of the coefficients of reduction 
of variance is presented in \callout{figure~\ref{f:dist}}. The weighted mean 
value $\bar{R}$ over 69 baselines is \mbox{ $0.97 \pm 0.04$ }. For the 
computation of the uncertainty of the mean value we used the variance of 
$R$ which can easily be derived from the results presented in the appendix 
of VDH:
\beq
       \var(R) = \Frac{\sigma_1^2 + \sigma_2^2 - \sigma_m^2}{2(N-1)\sigma_m^2 } 
\eeq{e:e4}
\BegCorr{C3}
where $\sigma_1^2$ and $\sigma_2^2$ are the mean squares of the baseline length
residuals of solutions B1 and B2 respectively, $\sigma_m^2$ is the mean square
of the modeled baseline changes with respect to the mean value computed over
the period of time of VLBI observation, $N$ is the number of observing 
sessions.
\EndCorr{C3}

\begin{figure}
   \ifjdraft \else 
        \hspace{-7mm}
        \mbox{ \epsfxsize=83mm \epsfclipon \epsffile{aplo_rdist.ps} }
   \fi
   \caption{Histogram of the distribution of the coefficients of reduction
            of variances.}
   \label{f:dist}
\end{figure}

\BegCorr{C4} 
\EndCorr{C4} 


   VDH analyzed 22 baselines for the period 1979 to 1992. They reported a much
lower value of the reduction of variance coefficient, 0.62, which means that 
38\% of the power of the signal in their series of atmospheric pressure 
loading is not present in the VLBI data. In order to check whether the 
differences may be due to changes in the quality of VLBI data collected after 
1992 we restricted our calculations to exactly the same set of observations and 
baselines used by VDH. We got $ 1.10 \pm 0.10 $ for the coefficient of 
reduction of variance for this case. It deviates at the 3$\sigma$ level from 
the value reported by VDH. The differences in our analysis technique of VLBI 
observations and the technique of VDH are not significant enough to explain 
this large discrepancy. So, the differences in the reduction of variance 
coefficients come from the differences between the series of the atmospheric 
pressure loading displacements. 
\BegCorr{C5} 
First, the predicted baseline loading effects of VDH and 
\citet {r:gpsatmloa} suffered from the problem that they have the wrong sign
for the horizontal deformation in the North direction (T. van Dam personal
communication, 2003). 
\EndCorr{C5} 
Second, we used the NCEP Reanalysis model, but VDH used
an older model: the National Meteorological Center operational model with 
a 12 hour resolution. Third, we used a land-sea mask with a resolution of 
$ \Deg{0}\!.25 \times \Deg{0}\!.25 $ --- ten times better then the resolution 
of the surface pressure model. As shown in section~\ref{s:err}, this 
difference alone can cause an error in the vertical displacements as large 
as 10\%. Fourth, we used a different numerical algorithm for computing the 
convolution integral.

\subsection{Application of the Pressure Loading Model to Data Reduction}

  In the past, several authors recommended finding the linear regression of 
the vertical atmospheric pressure loading and local surface pressure and 
using this simple model in operational data analysis. \citet{r:rab} warned that
since ``the magnitude of the displacements is critically dependent on the
spatial extension of the pressure distribution, [\ldots] there cannot be any
unique regression coefficient between local displacements and local air 
pressure changes which could be used to correct geodynamics measurements for 
air-pressure-induced surface displacements''. However, this simple model was
used in the 20th century. In order to evaluate the errors of this approach,
we computed coefficients of linear regression between the 22 year long 
time series of the displacements caused by atmospheric pressure loading and 
local pressure at the station obtained by interpolation of the NCEP surface 
pressure field. Using these regression coefficients we computed a synthetic 
time series of the pressure loading for each station and estimated the 
admittance factors of these time series in solution G7. The results are shown 
in table~\ref{t:gloadm}. We see that the simple regression model works 
surprisingly well for vertical components. This test shows that the amount 
of the power of the signal which is present in the model, but is not present 
in the data, is increased from 5\% to 12\% when a regression model is used.

   Currently, there is no need to resort to a simplified linear regression. 
Numerical weather models are available on-line promptly, and a computation 
of a 20 years long series of pressure loading for all VLBI and SLR sites takes 
only several days at a personal computer using the efficient algorithm 
presented in the Appendix. 

   Applying atmospheric pressure loading in a procedure of data reduction 
causes a small change in the resulting terrestrial reference frame: 
the maximum site position change among the stations which observed one year or
longer is 2 mm, the velocity change is typically below 0.1 mm/yr with the 
maximum change of 0.4 mm/yr, and the scale factor is increased by 
$0.05 \pm 0.02$ ppb. Taking into account the horizontal component of the 
displacement due to atmospheric pressure loading causes rms differences in the 
estimates of polar motion and UT1 at the level of 100 prad and differences 
in the estimates of nutation angles with an rms of 30 prad. The uncertainties 
of the EOP derived from processing daily VLBI sessions are currently at the 
level of 200--400 prad. Therefore, omission of the horizontal atmospheric 
pressure loading is currently not a significant source of noise in the 
estimates of the EOP.

\section{Conclusions}

   We found that vertical and horizontal displacements caused by atmospheric 
pressure loading currently can be computed with errors less than 15\% by 
convolution of the surface pressure field from the NCEP Reanalysis model with 
Green's functions. Our analysis of VLBI observations of 40 strong stations
for the time period from 1980 till 2002 demonstrates that on average only 
5\% of the power of the modeled vertical pressure loading signal, 16\% of 
horizontal signal and 3\% of the signal in baseline lengths is not found 
in VLBI data. Thus, all discrepancies between the observations and the model
can be explained by {\em known} deficiencies of the model.

   Admittance factors of the time series of pressure loading vertical 
displacements without the annual constituent do not deviate from unity by 
more than the formal uncertainty. It means that at the confidence level of 
5\% the modeled signal is completely recovered from the VLBI observables. This 
estimate sets the upper limit of possible errors in Green's functions: 
4\% for the radial Green's function and 12\% for the horizontal Green's 
function.

   This allows us to conclude that on average our model of the displacements
caused by atmospheric pressure loading is correct. At the same time for some 
stations the model does not agree with data perfectly. These stations are 
located either close to a coast, or in isolated islands, or in mountain 
regions. In the first case, poor modeling of the ocean response to atmospheric
pressure forcing becomes a significant factor; in the latter case the spatial 
resolution of the weather model, $ \Deg{2}\!.5 \times \Deg{2}\!.5 $, is too 
coarse to represent adequately the surface pressure field variations.

   We have detected for the first time the horizontal component of the 
atmospheric pressure loading signal. This signal has never been before taken 
into account in routine reduction of geodetic observations. If not modeled, 
it adds noise to the horizontal site position with an rms of 0.6 mm and to
the estimates of the EOP with an rms of 100 prad. According to the estimates
of the error budget, the residual site displacements due to atmospheric 
pressure loading which is not accounted for by our model, on average have 
an rms of 0.4 mm for the vertical component and 0.1 mm for the horizontal 
component.
    
   The model presented here shows a significant improvement with respect to the
previous models. The amount of power which is present in the model, but not 
found in the data is 38\% for the VDH model, 12\% when the linear regression 
approach is used, and 5\% for our model.

   We propose our model of the atmospheric pressure loading for use in 
routine processing of space geodesy observations. We have computed time series 
of the atmospheric pressure loading for all VLBI and SLR sites starting 
from May 1976. These time series are available on the Web at 
\mbox{\tt http://gemini.gsfc.nasa.gov/aplo} and are updated daily. Since 
December 2002 the contribution of the atmospheric pressure loading 
displacements is applied on a routine basis at the VLBI analysis center 
of the NASA Goddard Space Flight Center. 

  We expect that the new numerical weather models with higher spatial and 
temporal resolution will improve the agreement of pressure loading time 
series with observations in mountainous regions. Development of ocean models 
forced by atmospheric pressure and winds will improve the pressure 
loading estimates for coastal and island stations.

\appendix
\section{Algorithm for the Computation of Displacements Due to Atmospheric
pressure Loading}

  We represent each component of the displacement as a sum of the 
contribution of the convolution integral over the land and over the ocean:
\beq
        u(\vec{r},t) = u_{\sc l}(\vec{r},t) + \Delta \bar{P}_o(t) \, u_o
\eeq{e:a1}
where $ \Delta \bar{P}_o(t) $ is the uniform sea-floor pressure and 
$u_{\sc l}(\vec{r},t)$, $u_o(\vec{r}\,)$ are
\beq
  \begin{array}{rcl}
      u_{\sc l}(\vec{r},t) &=& \dss\sum_{i=1}^{n} \sum_{j=1}^{m} \, 
                           \Delta P(\vec{r}{\,'\!}_{ij},t) \, 
                           q(\vec{r},\vec{r}{\,'\!}_{ij}) \, \cos \phi_i \, 
                           \dint\limits_{\hbb\! cell_{ij}}       
                           G(\psi(\vec{r},\vec{r}{\,'}_{ij})) \, ds     
                           \vspace{0.5ex} \\
      u_o(\vec{r}\,) &=& \dss\sum_{i=1}^{n} \sum_{j=1}^{m} \, 
                           q(\vec{r},\vec{r}{\,'\!}_{ij}) \, \cos \phi_i \,
                           \dint\limits_{\hbb\! cell_{ij}} 
                           G(\psi(\vec{r},\vec{r}{\,'\!}_{ij})) \, ds 
   \end{array}
\eeq{e:a2}
 and index $i$ runs over latitude and index $j$ runs over longitude. Here 
we replaced the integration over the sphere with a sum of integrals over small
cells. $q=1$ for the vertical component.

  Green's functions have a singularity in 0, so care must be taken in using
numerical schemes for computing the convolution integral. Although the Green's
function cannot be represented analytically over the whole range of its 
argument, we can always find a good approximation over a small range. 
We approximate the function $ G(\psi) \cdot \psi $ by a polynomial of the 
third degree 
\mbox{$ \alpha + \beta \, \psi + \gamma \, \psi^2 + \delta \, \psi^3 $}.
In order to compute the integral \ref{e:a2} over the cell, we introduce 
a two-dimensional Cartesian coordinate system with the origin in the center 
of the cell and the axis $x$ towards east, the axis $y$ towards north. 
We neglect the Earth's curvature and consider the cell as a rectangle with 
borders [-a, a], [-b, b] on the $x$ and $y$ axes respectively. Then the 
integral of the Green's function over the cell with respect to a site with 
coordinates ($x_s$, $y_s$) is evaluated analytically:

\beq
   \begin{array}{ll}
      \ifjdraft
         \dss \dint\limits_{\hbb\!\! cell} G(\psi(x_s,y_s)) \, ds = & 
      \else
         \dss \dint\limits_{\hbb\!\! cell} G(\psi(x_s,y_s)) \, ds = & \\ & 
      \fi
          \dss \int\limits_{\hbb -b}^{b} \int\limits_{\hbb -a}^{a} \: 
                      \Biggl(
                             \Frac{\alpha}{\sqrt{x^2 + y^2}}  \: + \: 
                                   \beta                      \: + \: 
                                   \gamma \, \sqrt{x^2 + y^2} \: + \: 
                                   \delta (x^2 + y^2) 
                      \Biggr) \, dx \, dy =            \vex \vex \vex \vex \\ &
          \biggl( \alpha \, y_2 + \Frac{\gamma}{6} \, y^3_2 \biggr)
                  \ln \Frac{x_2 + z_{22}}{x_1 + z_{12}}       \: - \:
          \biggl( \alpha \, y_1 + \Frac{\gamma}{6} \, y^3_1 \biggr)
                  \ln \Frac{x_2 + z_{21}}{x_1 + z_{11}}       \: + \: \vex \\ &
          \biggl( \alpha \, x_2 + \Frac{\gamma}{6} \, x^3_2 \biggr)
                  \ln \Frac{y_2 + z_{22}}{y_1 + z_{21}}       \: - \:
          \biggl( \alpha \, x_1 + \Frac{\gamma}{6} \, x^3_1 \biggr)
                  \ln \Frac{y_2 + z_{12}}{y_1 + z_{11}}       \:\: +  \vex \\ &
          (y_2 - y_1) \, (x_2 - x_1) 
                \Biggr[ \beta + \Frac{\delta}{3} 
                        \biggl( 
                                 z^2_{11} + z^2_{22} + x_1 \, x_2 + y_1 \, y_2 
                        \biggr)
                \Biggr]                                       \:\: +  \vex \\ &
          \Frac{\gamma}{3} 
                \Biggr[
                    x_2 \, \biggl( y_2 \, z_{22} - y_1 \, z_{21} \biggr) \: - \:
                    x_1 \, \biggl( y_2 \, z_{12} - y_1 \, z_{11} \biggr)
                \Biggr]
      \vspace{3ex} \\ 
      \nntab{l}{
        \begin{array}{l @{\hspace{4em}} l}
           x_1 =  -a - x_s   &  x_2 = \hpm a - x_s  \vex \\
           y_1 =  -b - y_s   &  y_2 = \hpm b - y_s  \vex \\
           z_{11} = \sqrt{x_1^2 + y_1^2}  &  z_{12} = \sqrt{x_1^2 + y_2^2} 
                                                    \vex \\
           z_{21} = \sqrt{x_2^2 + y_1^2}  &  z_{22} = \sqrt{x_2^2 + y_2^2}  
        \end{array} 
      }
   \end{array}
\eeq{e:a3}
\par\medskip\par

  Coordinates $x_s, y_s$ are  computed as
\beq
        x_s = \vec{E}(\vec{r}{\,'\!}_{ij}) \cdot 
              \vec{T}(\vec{r}{\,'\!}_{ij},\vec{r}\,) \qquad\qquad
        y_s = \vec{N}(\vec{r}{\,'\!}_{ij}) \cdot 
              \vec{T}(\vec{r}{\,'\!}_{ij},\vec{r}\,)
\eeq{e:a4}
where $\vec{T}(\vec{r}{\,'\!}_{ij},\vec{r}\,)$ is 
\beq
     \vec{T}(\vec{r}{\,'\!}_{ij},\vec{r}\,) = \Frac
       {   \vec{r}{\,'\!}_{ij} \times [ \vec{r} \times \vec{r}{\,'\!}_{ij} ]   }
       { | \vec{r}{\,'\!}_{ij} \times [ \vec{r} \times \vec{r}{\,'\!}_{ij} ] | }
\eeq{e:a5}
and $\vec{E}(\vec{r}{\,'\!}_{ij}), \vec{N}(\vec{r}{\,'\!}_{ij})$ are unit 
vectors in east and north direction with respect to the center of the cell:
\beq
     \vec{E}(\vec{r}{\,'\!}_{ij}) = \left(
        \begin{array}{r}
           \sin \lambda'  \\
           \cos \lambda'  \\
	   0              \\
        \end{array}   
        \right)
        \qquad \qquad
     \vec{N}(\vec{r}{\,'\!}_{ij}) = \left(
        \begin{array}{r}
          \hpm \sin \phi' \cos \lambda'  \\
             - \sin \phi' \sin \lambda'  \\
	  \hpm \cos \phi'
        \end{array} 
        \right)
\eeq{e:a6}

  We found that when the coefficients $ \alpha(\psi) $, $ \beta(\psi) $, 
$ \gamma(\psi) $  and $ \delta(\psi) $ are computed with the step 0.002 rad 
over the range [0, 0.16] rad, and with the step 0.02 rad over the range 
[0.16, $\pi$], the error of the approximation of the integral~\ref{e:a3}
for a cell of size 0.044 rad ($\Deg{2}\!.5$) does not exceed 1\%. 
At large angular distances we can consider the Green's function to be constant 
over the cell. For an angular distance more than 0.16 rad, taking the Green's 
function out of the integral \ref{e:a2} and replacing it with the value 
at the angular distance between the site and the center of the cell causes
an error of less than 1\%.

  Two land-sea masks are used for practical computation: coarse with the 
resolution of the surface pressure grid, and fine. If the cell of the coarse 
land-sea mask is completely land or completely sea, this cell is used for 
computing the integral \ref{e:a3}. Otherwise, the coarse resolution cell is 
subdivided in smaller cells of the fine resolution grid, and the integral over 
each fine resolution cell is computed independently. The surface pressure is 
considered as defined at the corners of the coarse resolution cell. 
The pressure at the center of the cell is obtained by bi-linear interpolation. 
When $u_{\sc l}(\vec{r}\,)$ is computed, the cells which are over ocean are 
bypassed. Alternatively, the cells which are over land are bypassed when 
$u_o(\vec{r}\,)$ is computed.

  The computation of horizontal vectors is done separately for north and east 
components. The north and east components of the vector 
$\vec{q}(\vec{r},\vec{r}\,')$ are
\beq
    \vec{q}_{\sc n}(\vec{r},\vec{r}\,') = - \vec{T}(\vec{r},\vec{r}\,')
             \cdot \vec{N}(\vec{r}\,) \qquad\qquad
    \vec{q}_{\sc e}(\vec{r},\vec{r}\,') = - \vec{T}(\vec{r},\vec{r}\,')
             \cdot \vec{E}(\vec{r}\,) 
\eeq{e:a7}
  where $\vec{T}(\vec{r},\vec{r}\,')$ is defined in a way similar 
to \ref{e:a5}, but with the reverse order of arguments,
$\vec{E}(\vec{r}\,), \vec{N}(\vec{r}\,)$ are defined according to \ref{e:a6}, 
but are the unit north and east vectors for the site under consideration.

  Source code of the programs for computation of the displacements caused
by the atmospheric pressure loading is available on the Web 
at \mbox{\tt http://gemini.gsfc.nasa.gov/aplo}.

\begin{acknowledgments}

   We used NCEP Reanalysis data provided by the NOAA-CIRES Climate
Diagnostics Center, Boulder, Colorado, USA, from their Web site at
\mbox{\tt http://www.cdc.noaa.gov/}

  The authors have a pleasure to thank H. Goosse for providing outputs of CLIO 
model, R. Ray for providing his model of the atmospheric pressure tides
and T. van Dam for careful cross-checking our computations. We wish to 
thank D. Rowlands and A. Niell for valuable comments which helped to improve 
this paper. This work was done while Leonid Petrov worked for NVI, Inc. at 
Goddard Space Flight Center under NASA contract NAS5-01127 and Jean-Paul Boy 
was a NAS/NRC post-doctoral fellow at Goddard Space Flight Center.

\end{acknowledgments}

\end{article}

\begin{table}
  \small
  \caption{Statistics of Displacements Caused by Atmospheric Pressure 
           Loading and Statistics of the Differences in Time Series:
           NCEP Operational versus NCEP Reanalysis Dataset;
           CLIO Model versus IB Hypothesis.
           }
  \begin{tabular}{ l   @{\hspace{1.5em}} r    @{\hspace{1.5em}}r 
                       @{\hspace{3em}} rr   @{\hspace{3em}} 
                  rrrr @{\hspace{3em}} rrrr}
    \tableline 
      Station  & \ntab{c}{Lat}   & \ntab{l}{Long} & 
               \nntab{c}{Displacements \hspace{0.5em}\hpm} & 
               \nnnntab{c}{Operational/Reanalysis \hspace{2em}\hpm} & 
               \nnnntab{c}{CLIO/IB} 
               \vnx \\
               &                 &                 & 
               \nntab{c}{r.m.s.\hspace{2.5em}\hpm} & 
               \nntab{c}{rms of diff.} & 
               \nntab{c}{correlation \hspace{1.0em}\hpm} & 
               \nntab{c}{rms of diff.} & 
               \nntab{c}{correlation } 
               \vnx \\
             &      &     & vert. & horz. & vert.  & horz. & vert.  & 
              horz. & vert.  & horz. & vert.   & horz. \vnx \\
             & Deg  & Deg &  mm   & mm    &  \%    & \%    &        &
                    & \%     & \%    &         &            \\
    \tableline

    ALGOPARK  & $ \Deg{ 45}\!.8 $ & $ \Deg{281}\!.9 $ & 3.7 & 0.6 &
     4.0 &  5.7 & 0.96 & 0.99 &  4.7 & 11.0 &  0.02 &  0.00 \vnx \\
    BR-VLBA   & $ \Deg{ 47}\!.9 $ & $ \Deg{240}\!.3 $ & 3.2 & 0.6 &
     8.8 &  9.8 & 0.96 & 0.92 &  6.6 & 12.0 &  0.02 & -0.08 \vnx \\
    DSS45     & $ \Deg{-35}\!.2 $ & $ \Deg{149}\!.0 $ & 2.5 & 0.4 &
     5.8 &  9.1 & 0.97 & 0.90 &  9.0 & 13.1 &  0.03 & -0.13 \vnx \\
    DSS65     & $ \Deg{ 40}\!.2 $ & $ \Deg{355}\!.7 $ & 2.6 & 0.6 &
     8.2 & 12.7 & 0.97 & 0.77 &  7.6 & 15.0 &  0.08 &  0.02 \vnx \\
    FD-VLBA   & $ \Deg{ 30}\!.5 $ & $ \Deg{256}\!.1 $ & 2.4 & 0.4 &
    10.4 &  9.4 & 0.93 & 0.95 &  9.2 & 19.1 & -0.05 & -0.07 \vnx \\
    FORTLEZA  & $ \Deg{ -3}\!.9 $ & $ \Deg{321}\!.6 $ & 1.0 & 0.3 &
    15.5 & 12.5 & 0.94 & 0.86 & 25.4 & 23.3 &  0.35 & -0.07 \vnx \\
    GILCREEK  & $ \Deg{ 64}\!.8 $ & $ \Deg{212}\!.5 $ & 4.7 & 0.6 &
     7.8 & 12.3 & 0.99 & 0.95 &  2.7 & 16.2 &  0.12 &  0.06 \vnx \\
    HARTRAO   & $ \Deg{-25}\!.7 $ & $ \Deg{ 27}\!.7 $ & 2.3 & 0.2 &
     7.8 & 18.3 & 0.88 & 0.91 & 10.3 & 16.4 & -0.01 &  0.04 \vnx \\
    HATCREEK  & $ \Deg{ 40}\!.6 $ & $ \Deg{238}\!.6 $ & 2.0 & 0.5 &
    10.3 &  9.3 & 0.97 & 0.94 &  9.4 & 14.4 & -0.04 & -0.10 \vnx \\
    HAYSTACK  & $ \Deg{ 42}\!.4 $ & $ \Deg{288}\!.5 $ & 2.8 & 0.6 &
     4.5 &  4.8 & 0.92 & 0.94 & 11.2 & 14.6 & -0.01 &  0.06 \vnx \\
    HN-VLBA   & $ \Deg{ 42}\!.7 $ & $ \Deg{288}\!.0 $ & 3.0 & 0.6 &
     4.5 &  5.2 & 0.92 & 0.96 &  6.6 & 19.6 &  0.02 &  0.04 \vnx \\
    HOBART26  & $ \Deg{-42}\!.6 $ & $ \Deg{147}\!.4 $ & 1.6 & 0.4 &
     4.9 &  8.1 & 0.93 & 0.95 & 12.0 & 33.4 & -0.05 & -0.07 \vnx \\
    HRAS  085 & $ \Deg{ 30}\!.5 $ & $ \Deg{256}\!.0 $ & 2.4 & 0.4 &
    10.4 &  9.4 & 0.93 & 0.95 &  9.2 & 19.1 &  0.01 & -0.07 \vnx \\
    KASHIMA   & $ \Deg{ 35}\!.8 $ & $ \Deg{140}\!.7 $ & 1.2 & 0.6 &
     7.4 &  6.0 & 0.96 & 0.93 & 17.3 & 30.2 &  0.11 &  0.00 \vnx \\
    KASHIM34  & $ \Deg{ 35}\!.8 $ & $ \Deg{140}\!.7 $ & 1.2 & 0.6 &
     7.5 &  6.0 & 0.96 & 0.93 & 17.4 & 33.8 &  0.10 &  0.01 \vnx \\
    KAUAI     & $ \Deg{ 22}\!.0 $ & $ \Deg{200}\!.3 $ & 0.5 & 0.2 &
    11.7 & 10.5 & 0.99 & 0.96 & 52.4 & 87.3 & -0.09 &  0.06 \vnx \\
    KOKEE     & $ \Deg{ 22}\!.0 $ & $ \Deg{200}\!.3 $ & 0.5 & 0.2 &
    11.7 & 10.5 & 0.99 & 0.96 & 52.4 & 89.9 & -0.09 &  0.06 \vnx \\
    KP-VLBA   & $ \Deg{ 31}\!.8 $ & $ \Deg{248}\!.4 $ & 1.9 & 0.5 &
    13.8 & 10.9 & 0.87 & 0.93 & 12.1 & 44.6 & -0.09 & -0.12 \vnx \\
    LA-VLBA   & $ \Deg{ 35}\!.6 $ & $ \Deg{253}\!.7 $ & 2.5 & 0.5 &
     8.1 &  9.6 & 0.99 & 0.95 &  8.6 & 44.1 & -0.07 & -0.06 \vnx \\
    MATERA    & $ \Deg{ 40}\!.5 $ & $ \Deg{ 16}\!.7 $ & 2.3 & 0.8 &
     9.9 &  8.7 & 0.92 & 0.75 &  9.1 & 40.5 & -0.03 & -0.03 \vnx \\
    MEDICINA  & $ \Deg{ 44}\!.3 $ & $ \Deg{ 11}\!.6 $ & 3.3 & 0.9 &
     7.0 &  9.3 & 0.86 & 0.75 &  6.9 & 38.0 & -0.01 & -0.01 \vnx \\
    MK-VLBA   & $ \Deg{ 19}\!.7 $ & $ \Deg{204}\!.5 $ & 0.5 & 0.2 &
    12.3 & 10.6 & 0.74 & 0.76 & 53.0 & 37.5 & -0.12 &  0.03 \vnx \\
    MOJAVE12  & $ \Deg{ 35}\!.2 $ & $ \Deg{243}\!.1 $ & 1.9 & 0.5 &
    12.4 & 11.7 & 0.85 & 0.89 & 11.4 & 60.6 & -0.07 & -0.11 \vnx \\
    NL-VLBA   & $ \Deg{ 41}\!.6 $ & $ \Deg{268}\!.4 $ & 3.6 & 0.7 &
     4.3 &  6.1 & 0.96 & 0.94 &  5.7 & 44.5 &  0.03 & -0.02 \vnx \\
    NOTO      & $ \Deg{ 36}\!.7 $ & $ \Deg{ 15}\!.0 $ & 1.4 & 0.7 &
    12.9 & 11.2 & 0.98 & 0.77 & 13.6 & 62.2 & -0.02 & -0.03 \vnx \\
    NRAO20    & $ \Deg{ 38}\!.2 $ & $ \Deg{280}\!.2 $ & 3.0 & 0.5 &
     6.5 &  7.1 & 0.92 & 0.97 &  6.8 & 58.1 &  0.04 &  0.03 \vnx \\
    NRAO85  3 & $ \Deg{ 38}\!.2 $ & $ \Deg{280}\!.2 $ & 3.0 & 0.5 &
     6.5 &  7.1 & 0.92 & 0.97 &  6.8 & 60.6 &  0.04 &  0.03 \vnx \\
    NYALES20  & $ \Deg{ 78}\!.9 $ & $ \Deg{ 11}\!.9 $ & 1.8 & 0.7 &
     6.7 &  7.0 & 0.60 & 0.83 &  5.8 & 62.6 &  0.03 & -0.01 \vnx \\
    ONSALA60  & $ \Deg{ 57}\!.2 $ & $ \Deg{ 11}\!.9 $ & 4.7 & 1.1 &
     2.8 &  4.5 & 0.60 & 0.77 &  4.7 & 27.1 & -0.03 & -0.00 \vnx \\
    OV-VLBA   & $ \Deg{ 37}\!.0 $ & $ \Deg{241}\!.7 $ & 2.0 & 0.5 &
    12.7 & 10.9 & 0.88 & 0.91 & 10.1 & 28.6 & -0.06 & -0.11 \vnx \\
    OVRO  130 & $ \Deg{ 37}\!.0 $ & $ \Deg{241}\!.7 $ & 2.0 & 0.5 &
    12.9 & 10.9 & 0.87 & 0.91 & 10.1 & 33.8 & -0.06 & -0.11 \vnx \\
    PIETOWN   & $ \Deg{ 34}\!.1 $ & $ \Deg{251}\!.9 $ & 2.3 & 0.4 &
     9.8 & 10.4 & 0.90 & 0.93 &  8.8 & 40.3 & -0.07 & -0.08 \vnx \\
    RICHMOND  & $ \Deg{ 25}\!.5 $ & $ \Deg{279}\!.6 $ & 0.9 & 0.4 &
     7.9 &  6.1 & 0.81 & 0.86 & 26.5 & 47.6 & -0.03 &  0.02 \vnx \\
    SANTIA12  & $ \Deg{-33}\!.0 $ & $ \Deg{289}\!.3 $ & 1.4 & 0.4 &
    15.7 & 16.0 & 0.52 & 0.73 & 15.7 & 44.6 &  0.18 & -0.04 \vnx \\
    SC-VLBA   & $ \Deg{ 17}\!.6 $ & $ \Deg{295}\!.4 $ & 0.5 & 0.3 &
    15.7 & 10.0 & 0.89 & 0.85 & 55.9 & 44.6 &  0.10 &  0.00 \vnx \\
    SESHAN25  & $ \Deg{ 30}\!.9 $ & $ \Deg{121}\!.2 $ & 3.5 & 0.9 &
     7.6 &  7.8 & 0.77 & 0.85 &  7.1 & 34.0 &  0.03 &  0.01 \vnx \\
    TSUKUB32  & $ \Deg{ 35}\!.9 $ & $ \Deg{140}\!.1 $ & 1.4 & 0.5 &
     7.0 &  6.9 & 0.95 & 0.90 & 14.4 & 31.2 &  0.11 &  0.01 \vnx \\
    VNDNBERG  & $ \Deg{ 34}\!.4 $ & $ \Deg{239}\!.4 $ & 1.0 & 0.4 &
     8.3 &  8.8 & 0.87 & 0.91 & 19.8 & 30.2 & -0.08 & -0.12 \vnx \\
    WESTFORD  & $ \Deg{ 42}\!.4 $ & $ \Deg{288}\!.5 $ & 2.8 & 0.6 &
     4.5 &  4.8 & 0.92 & 0.94 &  7.7 & 31.7 &  0.02 &  0.04 \vnx \\
    WETTZELL  & $ \Deg{ 49}\!.0 $ & $ \Deg{ 12}\!.9 $ & 4.9 & 0.9 &
     4.2 &  6.4 & 0.77 & 0.80 & 16.9 & 28.3 & -0.03 &  0.04 \\
    \tableline \vspace{0.01ex}     
    Mean     &  &      & 2.6  & 0.6  &  8.7 &  9.1  & 0.89  & 0.89 &
     15.2 & 39.1 & 0.01 & -0.02 \\
    \tableline
  \end{tabular}
  \label{t:statist}
\end{table}

\ifjdraft  \newpage \fi

\begin{table}
   \caption{The Global Budget of the Errors of Computation of the 
            Atmospheric Pressure Loading Displacements.}
   \begin{tabular}{l @{\hspace{1em}} r @{\hspace{1.0em}} r }
      \tableline
         Error source                          & rms       & correlation \\ 
      \tableline
	Numerical evaluation of the integral &     1\%   & 0.0?      \hsp     \\
	Green's function                     & $ < 2\% $ & $\pm$ 1.0 \hsp\hpq \\
        Land sea mask                        & $ < 5\% $ & 0.7       \hsp\hpq \\
        Surface pressure field               &    10\%   & 0.9       \hsp\hpq \\
        Ocean response                       &    10\%   & 0.0       \hsp\hpq \\
      \tableline \vspace{-2ex}  \\
         Total                                 &    15\%   \\ 
      \tableline
   \end{tabular}
   \ifjdraft  \vspace{100mm} \fi
   \label{t:errbud}
\end{table}

\ifjdraft  \newpage \par\hpm\par\vspace{30mm} \fi

\begin{table}
   \caption{Global Admittance Factors.}
   \begin{tabular}{l @{\hspace{0.5em}} r @{\hspace{2em}} r @{\hspace{2em}} r}
       \tableline
          Solution & \ntab{c}{Up} & \ntab{c}{East} & \ntab{c}{North} \\
       \tableline
         G1 & 0.95 $\pm$ 0.02  &   1.17 $\pm$ 0.06     &  0.83 $\pm$ 0.07    \\
         G5 & 0.46 $\pm$ 0.09  &   1.08 $\pm$ 0.26     & -0.89 $\pm$ 0.26    \\
         G6 & 0.98 $\pm$ 0.02  &   1.20 $\pm$ 0.06     &  1.02 $\pm$ 0.07    \\
         G7 & 0.88 $\pm$ 0.02  &                       &                     \\
       \tableline
   \end{tabular}
   \label{t:gloadm}
   \ifjdraft  \vspace{100mm} \fi
\end{table}

\ifjdraft  \newpage \par\hpm\par\vspace{30mm} \fi

\begin{table}
   \caption{$\cal{P}$ of the Estimates of Residual Amplitudes of Harmonic
            Site Position Variations without Applying the Model of Atmospheric
            Pressure Loading and after Applying the Model.}
   \par\vspace{4ex}\par
   \begin{tabular}{l @{\hspace{3em}} c c @{\hspace{3em}} c c}
      \tableline
      Wave         & \nntab{l}{Without model} & \nntab{c}{With model} \\
                   &  vert   &  horz          &  vert  & horz   \\
      \tableline
      semi-diurnal &   2.77  &  1.73          &  2.35  &  1.58  \\
      diurnal      &   4.26  &  2.31          &  4.33  &  2.25  \\
      semi-annual  &   2.77  &  1.07          &  2.31  &  1.10  \\
      annual       &   5.18  &  2.45          &  6.00  &  2.45  \\
      \tableline
      \label{t:chi}
  \end{tabular}
\end{table}

\ifjdraft  \newpage \fi

\begin{table}
   \caption{Admittance Factors from G2 Solution.}
   \begin{tabular}{l @{\hspace{2em}}  l @{\hspace{2em}}  l @{\hspace{2em}} l}
      \tableline
          Station & \ntab{c}{Up} & \ntab{c}{East} & \ntab{c}{North} \\
      \tableline
   ALGOPARK    & \hpm 1.14 $\pm$ 0.09  & 1.03 $\pm$ 0.18 & 0.82 $\pm $ 0.21 \\
   BR-VLBA     & \hpm 0.75 $\pm$ 0.07  & 1.11 $\pm$ 0.13 & 0.42 $\pm $ 0.12 \\
   DSS45       & \hpm 0.85 $\pm$ 0.40  & & \\
   DSS65       & \hpm 2.30 $\pm$ 0.34  & & \\
   FD-VLBA     & \hpm 1.35 $\pm$ 0.07  & 2.44 $\pm$ 0.18 & 1.03 $\pm $ 0.13 \\
   FORTLEZA    & \hpm 1.08 $\pm$ 0.46  & & \\
   GILCREEK    & \hpm 0.94 $\pm$ 0.03  & 1.18 $\pm$ 0.11 & 0.35 $\pm $ 0.15 \\
   HARTRAO     & \hpm 1.53 $\pm$ 0.33  & & \\
   HAYSTACK    & \hpm 1.00 $\pm$ 0.36  & & \\
   HN-VLBA     & \hpm 0.12 $\pm$ 0.11  & 1.09 $\pm$ 0.16 & 1.03 $\pm $ 0.18 \\
   HRAS 085    & \hpm 1.84 $\pm$ 0.39  & & \\
   KOKEE       &     -0.95 $\pm$ 0.37  & & \\
   KP-VLBA     & \hpm 0.92 $\pm$ 0.11  & 0.33 $\pm$ 0.19 & 0.07 $\pm $ 0.14 \\
   LA-VLBA     & \hpm 0.36 $\pm$ 0.06  & 1.33 $\pm$ 0.15 & 0.68 $\pm $ 0.12 \\
   MATERA      & \hpm 0.79 $\pm$ 0.15  & 0.12 $\pm$ 0.28 & 1.17 $\pm $ 0.22 \\
   MEDICINA    & \hpm 0.67 $\pm$ 0.12  & 0.15 $\pm$ 0.25 & 1.20 $\pm $ 0.19 \\
   MOJAVE12    & \hpm 0.89 $\pm$ 0.24  & 1.93 $\pm$ 0.39 & 0.23 $\pm $ 0.38 \\
   NL-VLBA     & \hpm 0.91 $\pm$ 0.05  & 1.80 $\pm$ 0.10 & 0.61 $\pm $ 0.11 \\
   NRAO20      & \hpm 1.41 $\pm$ 0.08  & 1.99 $\pm$ 0.23 & 0.07 $\pm $ 0.21 \\
   NRAO85 3    & \hpm 1.56 $\pm$ 0.16  & 1.61 $\pm$ 0.42 & 0.67 $\pm $ 0.35 \\
   NYALES20    & \hpm 0.69 $\pm$ 0.12  & 0.60 $\pm$ 0.14 & 1.56 $\pm $ 0.18 \\
   ONSALA60    & \hpm 0.27 $\pm$ 0.08  & 1.03 $\pm$ 0.12 & 1.59 $\pm $ 0.13 \\
   OV-VLBA     &     -0.73 $\pm$ 0.11  & 0.94 $\pm$ 0.16 & 0.34 $\pm $ 0.15 \\
   PIETOWN     & \hpm 0.12 $\pm$ 0.07  & 1.18 $\pm$ 0.17 & 0.35 $\pm $ 0.13 \\
   SESHAN25    & \hpm 4.03 $\pm$ 0.39  & & \\
   TSUKUB32    & \hpm 3.91 $\pm$ 0.38  & & \\
   WESTFORD    & \hpm 1.22 $\pm$ 0.06  & 0.68 $\pm$ 0.11 & 0.06 $\pm $ 0.13 \\
   WETTZELL    & \hpm 1.14 $\pm$ 0.04  & 0.55 $\pm$ 0.12 & 0.91 $\pm $ 0.10 \\
      \tableline
   \end{tabular}
   \label{t:g2}
\end{table}

\ifjdraft

\newpage \par\hpm\par\vspace{5mm} \par

\vbox{ 
       \mbox{ \hspace{1mm} a) \hspace{81mm} b) } \par\medskip\par
       \mbox{ \epsfxsize=75mm \epsfclipon \epsffile{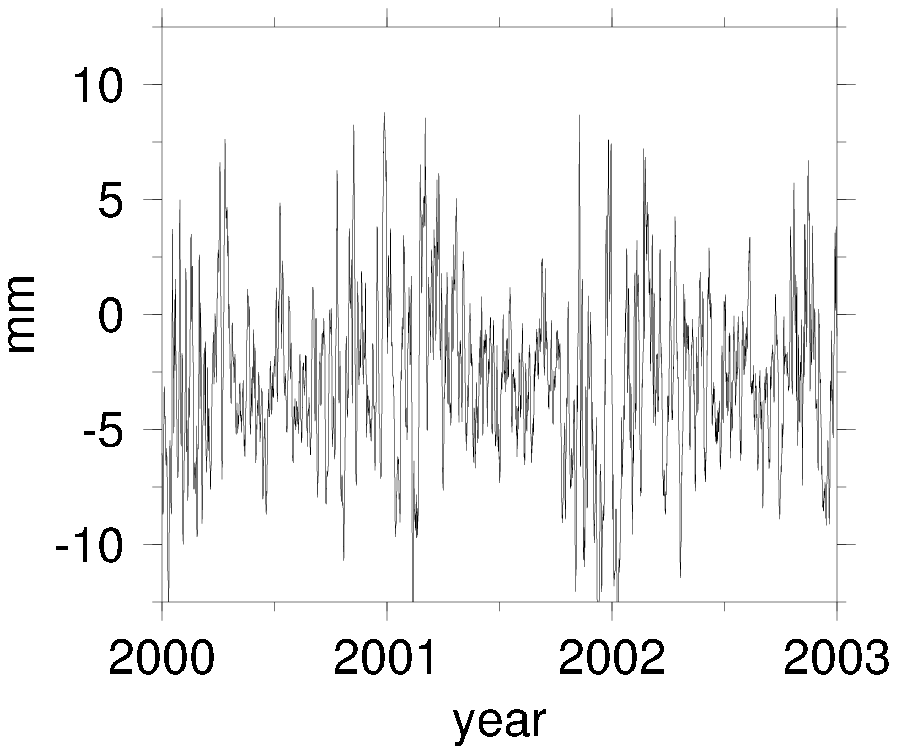} 
              \hspace{4mm}
              \epsfxsize=77mm \epsfclipon \epsffile{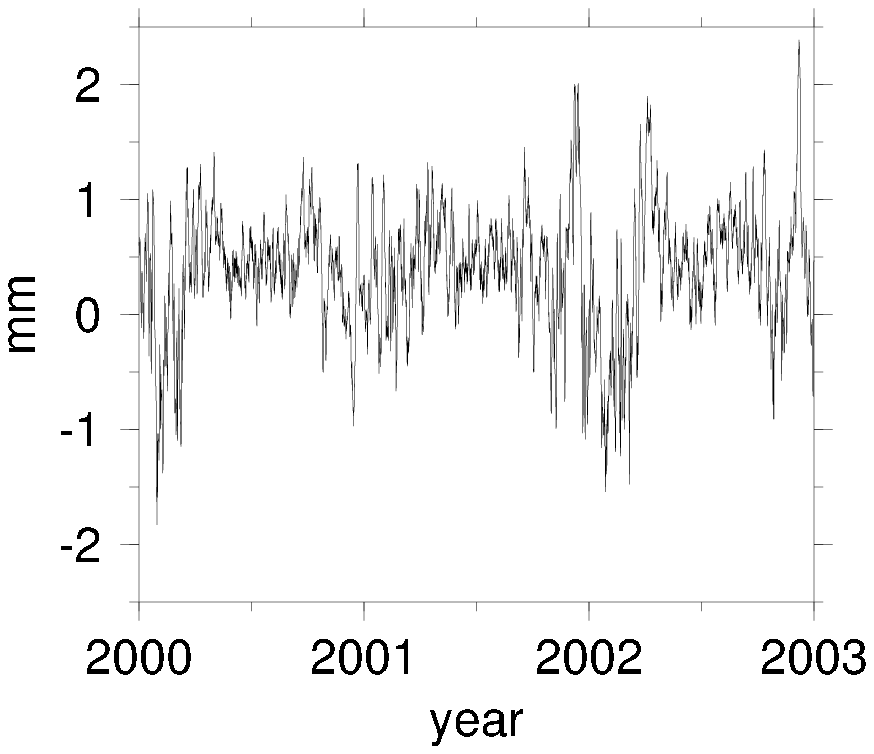} 
            }
       \par\vspace{2mm}\par
       \mbox{ \hspace{0mm} c) \hspace{78mm} d) } \par\medskip\par
       \mbox{ \epsfxsize=74mm \epsfclipon \epsffile{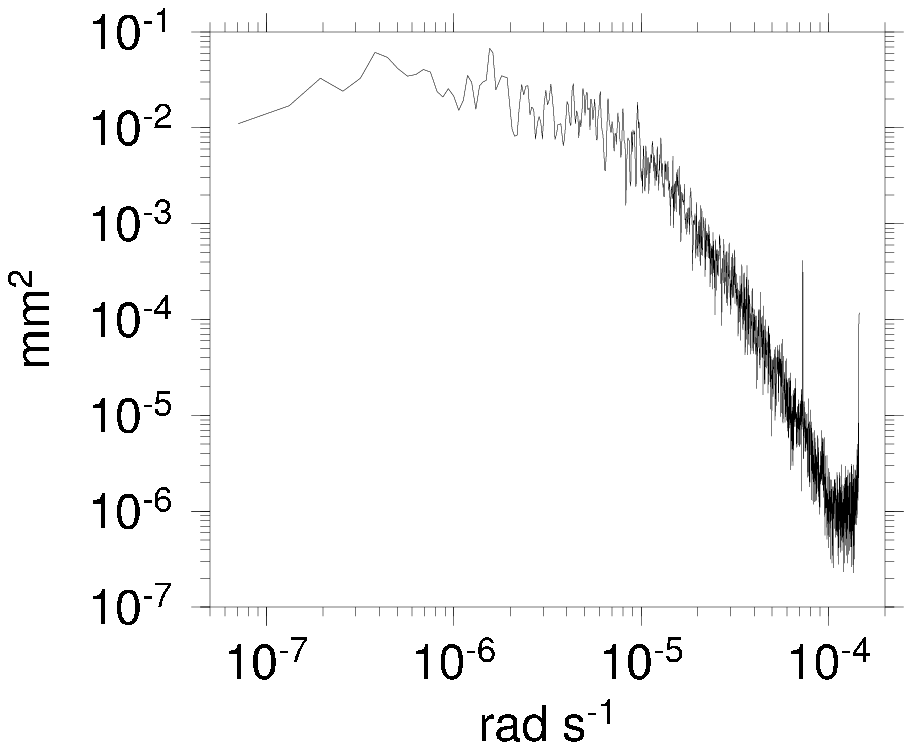}
              \hspace{4mm}
              \epsfxsize=74mm \epsfclipon \epsffile{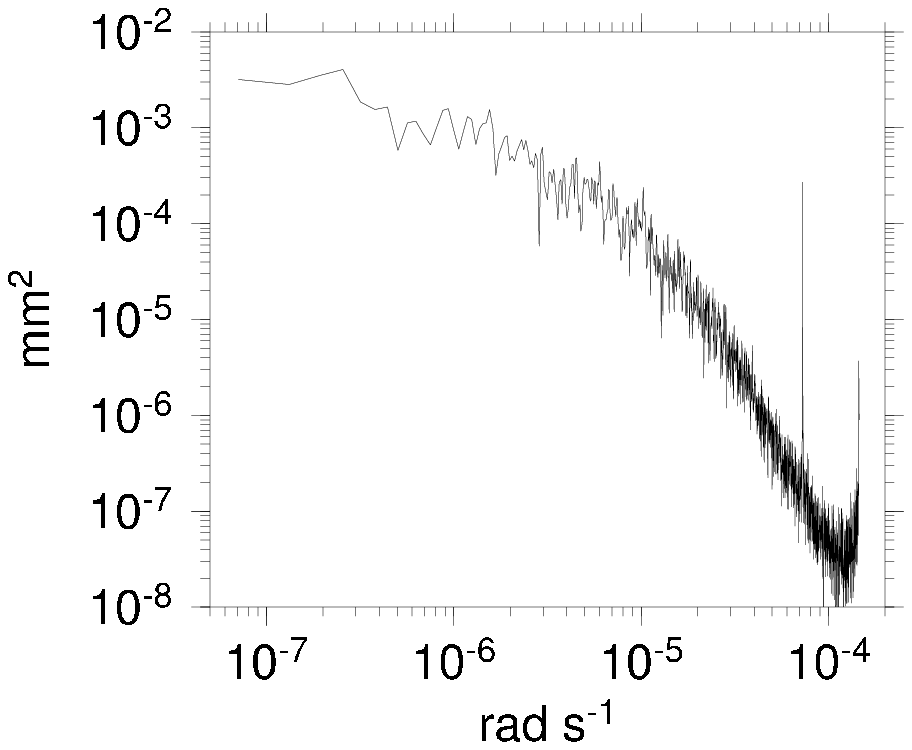} 
            }
       \par\bigskip\par
       \par Fig.~\ref{f:wettzell} \qquad \theauthors
     }
\newpage \par\hpm\par\vspace{5mm} \par

\vbox{ 
       \mbox{ \hspace{1mm} a) \hspace{81mm} b) } \par\medskip\par
       \mbox{ \epsfxsize=75mm \epsfclipon \epsffile{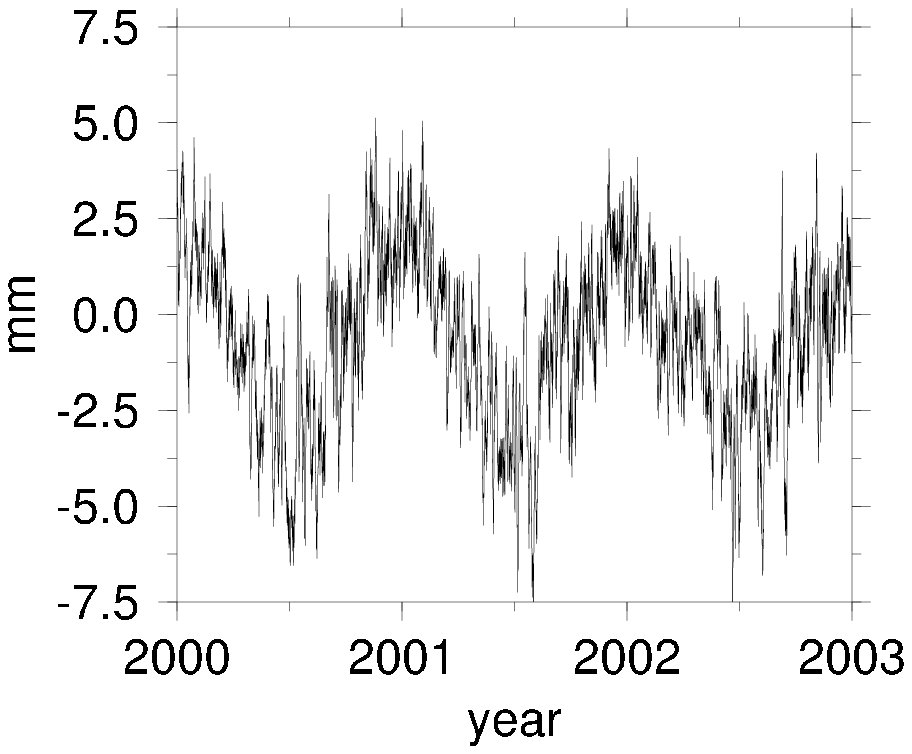} 
              \hspace{4mm}
              \epsfxsize=77mm \epsfclipon \epsffile{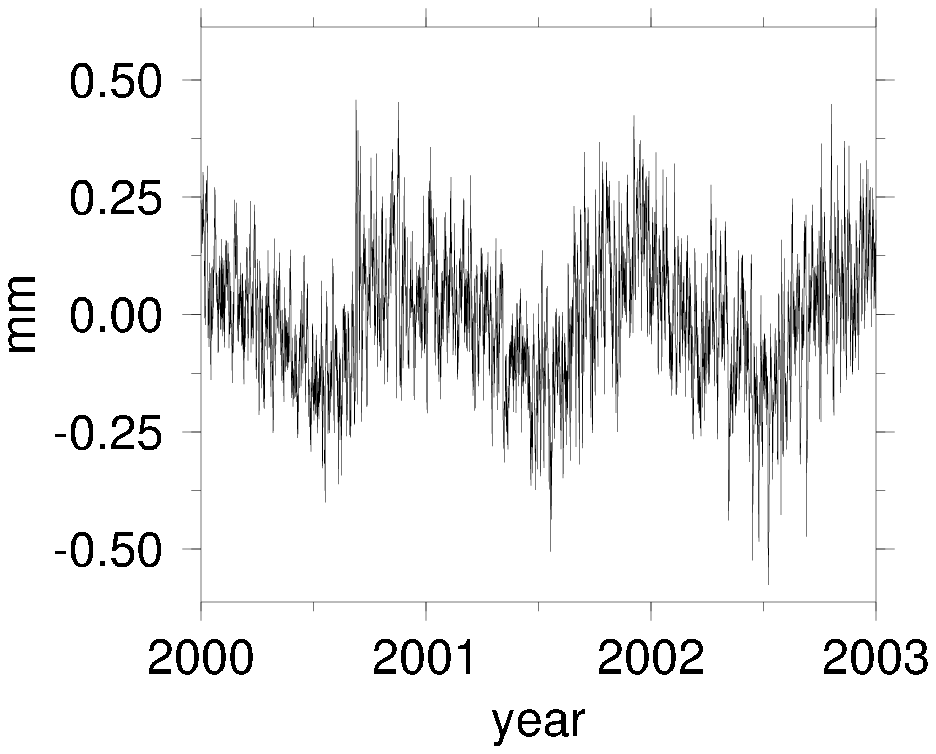} 
            }
       \par\vspace{2mm}\par
       \mbox{ \hspace{0mm} c) \hspace{78mm} d) } \par\medskip\par
       \mbox{ \epsfxsize=74mm \epsfclipon \epsffile{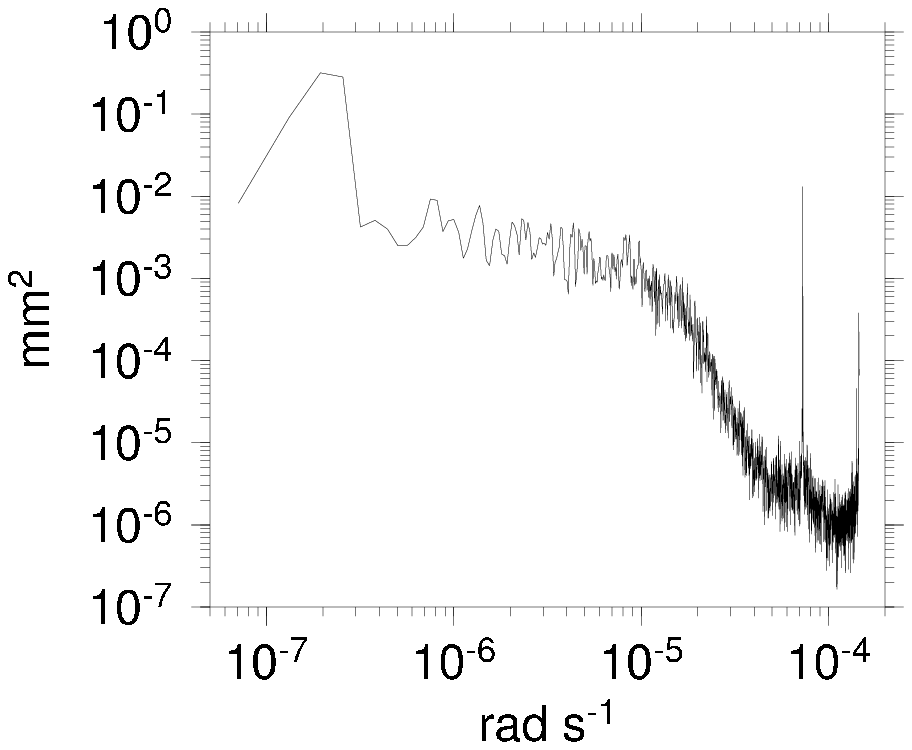}
              \hspace{4mm}
              \epsfxsize=74mm \epsfclipon \epsffile{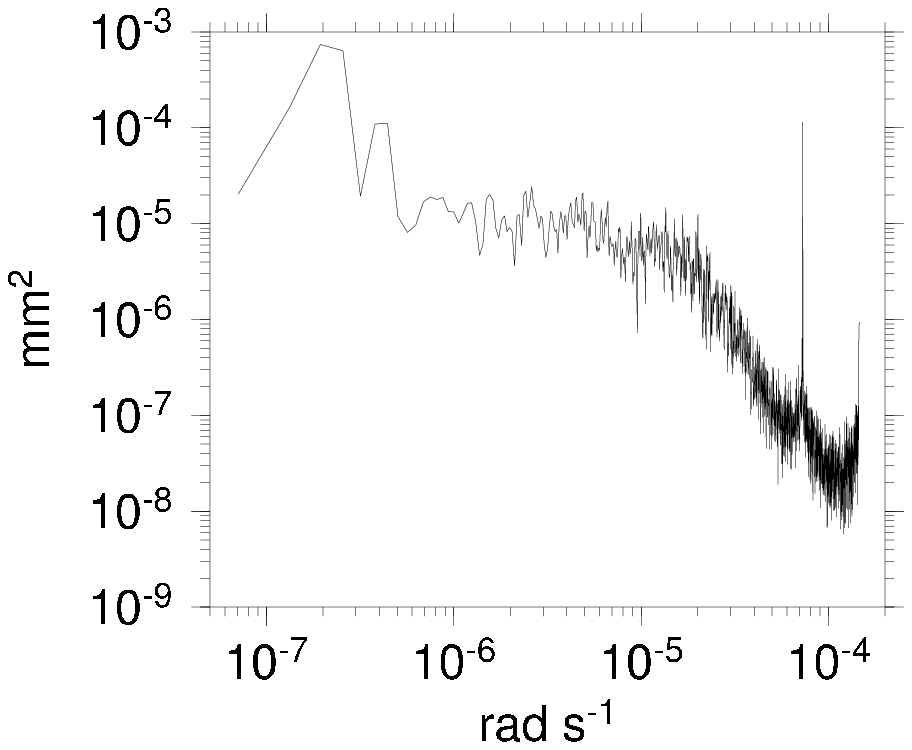} 
            }
       \par\bigskip\par
       \par Fig.~\ref{f:hartrao} \qquad \theauthors
     }

\vbox{ 
        \mbox{ \epsfxsize=73mm \epsfclipon \epsffile{aplo_autocorr.ps} }
        \par\vspace{-6mm}\par\noindent \hspace{54mm} Time (days)
        \par\bigskip\par
        Fig.~\ref{f:tcorr} \qquad \theauthors
     }

\newpage \par\hpm\par\vspace{50mm} \par

\vbox{ 
        \mbox{ \epsfxsize=73mm \epsfclipon \epsffile{aplo_spat_corr.ps} }
        \par\vspace{-6mm}\par\noindent \hspace{51mm} Distance (km) 
        \par\bigskip\par
        Fig.~\ref{f:scorr} \qquad \theauthors
     }

\newpage \par\hpm\par\vspace{50mm} \par

\vbox{ 
        \mbox{ \epsfxsize=75mm \epsfclipon \epsffile{aplo_map.ps} }
        \par Fig.~\ref{f:map} \qquad \theauthors
     }

\newpage \par\hpm\par\vspace{50mm} \par

\vbox{ 
        \mbox{ \epsfxsize=74mm \epsfclipon \epsffile{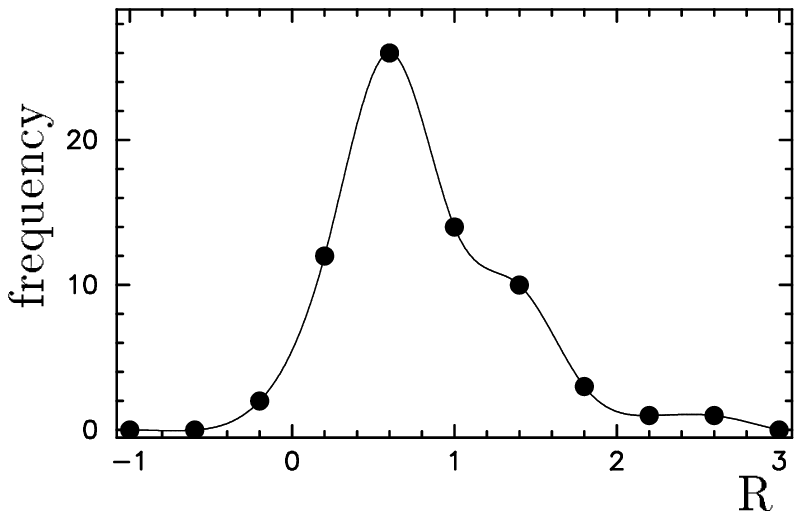} }
        \par\bigskip\par
        \par Fig.~\ref{f:dist} \qquad \theauthors
     }

%

\fi  


\begin{thebibliography}{33}

\bibitem[{\it Darwin}(1882)]{r:darwin}
   Darwin,~G.H., On variations in the vertical due to elasticity of the 
Earth's surface, {\it Phil. Mag.}, Ser.~5, {\it 14}, N.~90, 409--427, 1882.

\bibitem[{\it de Viron et al.}(2002)]{r:deviron}
   de Viron,~O., H.~Goosse, C.~Bizouard, and S.~Lambert,
High frequency non-tidal of the ocean on the Earth's rotation,
{\it EGS 27th General Assembly}, Nice, France, 2002.

\bibitem[{\it Dong et al.}(2002)]{r:dong}
   Dong,~D., P.~Fang, Y.~Bock, M.~K.~Cheng, and S.~Miyazaki, Anatomy of 
apparent seasonal variations from GPS derived site position time series, \jgr 
{\it 107}(B4), 10.1029/2001JB000573, 2002.

\bibitem[{\it Dziewonski and Andersen}(1981)]{r:PREM}
   Dziewonski,~A.M. and D.L.~Anderson, Preliminary Reference Earth Model,
{\it Phys. Earth Planet. Inter.} {\it 25}, 297--356, 1981.

\bibitem[{\it Farrell}(1972)]{r:farrell}
  Farrell,~W.E, Deformation of the Earth by Surface Loads, {\it Rev.\ Geophys.\
and Spac.\ Phys.}, {\it 10}(3), 751--797, 1972.

\bibitem[{\it Goosse and Fichelet}(1999)]{r:clio}
  Goosse,~H. and T.~Fichelet, Importance of ice-ocean interactions for the
ocean circulation: a model study, \jgr {\it 104}, 23,337--23,355, 1999. 

\bibitem[{\it Kalnay et al.}(1996)]{r:ncep}
   Kalnay~E., M.~Kanamitsu, R.~Kistler, W.~Collins, D.~Deaven, 
L.~Gandin, M.~Iredell, S.~Saha, G.~White, J.~Woollen, Y.~Zhu, 
A.~Leetma, R.~Reynolds, M.~Chelliah, W.~Ebisuzaki, W.Higgins, J.~Janowiak, 
K.~C.~Mo, C.~Ropelewski, J.~Wang, R.~Jenne and D.~Joseph, The NCEP/NCAR 
40-Year Reanalysis Project, {\it Bullet. Amer. Meteorol. Soc.}, {\it 77}, 
437--471, 1996.

\bibitem[{\it Lef\`{e}vre et al.}(2000)]{r:yellowsea}
   Lef\`{e}vre,~F., C.~Le Provost and F.H.~Lyard, 
How can we improve a global ocean tide model at a regional scale?
A test on the Yellow Sea and the East China Sea, \jgr {\it 105},
8707--8725, 2000.

\bibitem[{\it Lef\`{e}vre et al.}(2002)]{r:mask}
   Lef\`{e}vre,~F., F.H.~Lyard, C.~Le Provost and E.J.O.~Schrama,
FES99: a global tide finite element solution assimilating tide gauge and
altimetric information, {\it J. Atmos. Oceanic Technol.}, {\it 19},
1345--1356, 2002.

\bibitem[{\it MacMillan and Gipson}(1994)]{r:danatmload} 
   MacMillan,~D.S. and J.M.~Gipson, Atmospheric pressure loading parameters 
from very long baseline interferometric observations, \jgr {\it 99}(B9),
18,081--18,087, 1994.

\bibitem[{\it Magie}(1963)]{r:torr}
   Magie,~W.F., A source book in physics. Cambridge, Harvard University Press, 
1963.

\bibitem[{\it Manabe et al.}(1991)]{r:manabe} 
   Manabe,~S., T.~Sato, S.~Sakai, K.~Yokoyama, Atmospheric load effect on
VLBI observations, {Proc. of the AGU Chapman conference on geodetic VLBI:
Monitoring global change}, NOAA TR NOS~437, NGS~49, Washington~D.C., 111--122,
1991.

\bibitem[{\it Matsumoto et al.}(2000)]{r:nao99} 
   Matsumoto,~K., T.~Takanezawa, and M.~Ooe, Ocean Tide Models Developed by 
Assimilating TOPEX/POSEIDON Altimeter Data into Hydrodynamical Model: A Global 
Model and a Regional Model Around Japan, {\it Journal of Oceanography}, 
{\it 56}, 567--581, 2000.

\bibitem[{\it Milly and Shmakin}(2002a)]{r:millya} 
   Milly,~P.C.D. and A.B.~Shmakin, Global modeling of land water
and energy balances. Part I: the Land Dynamics (LaD) model, 
{\it J. Hydrometeor.}, {\it 3}, pp.~283--299, 2002a.

\bibitem[{\it Milly and Shmakin}(2002b)]{r:millyb} 
   Milly,~P.C.D. and A.B.~Shmakin, Global modeling of land water and energy 
balances. Part II: land-characteristic contributions to spatial variability, 
{\it J. Hydrometeor.}, {\it 3}, pp.~301--310, 2002b.

\bibitem[{\it McCarthy}(1996)]{r:iers96} 
   McCarthy,~D.D., IERS Conventions 1996, {\it IERS Technical Note}, No.~21, 
Paris, 1996.

\bibitem[{\it Niell}(1996)]{r:niell} 
   Niell,~A.E., Global mapping functions for the atmosphere delay at radio
wavelengths, \jgr {\it 100}, 3227--3246, 1996.

\bibitem[{\it Nothnagel et al.}(1995)]{r:termexp} 
   Nothnagel,~A., M.~Pilhatsch, and R.~Haas, Investigations of thermal height 
changes of geodetic VLBI radio telescopes, {\it Proceedings of the 10th 
Working Meeting on European VLBI for Geodesy and Astrometry}, edited by 
R.~Lanotte and G.~Bianco, Agenzia Spatiale Italiana, Matera, 121--133, 
1995.

\bibitem[{\it Okubo and Tsuji}(2001)]{r:okubo}
   Okubo,~S. and D.~Tsuji, Complex Green's function for diurnal/semidiurnal 
loading problems, {\it J. Geod. Soc. Japan}, {\it 47}, 225--230, 2001.

\bibitem[{\it Pagiatakis}(1990)]{r:pagiatakis}
   Pagiatakis,~S. D., The response of a realistic Earth to ocean tide loadings,
{\it Geophys. J. Int.}, {\it 103}, 541--560, 1990.

\bibitem[{\it Petrov}(2000a)]{r:peteop} 
   Petrov,~L., Estimation of EOP from VLBI: direct approach,
{\it Proc. of the IAU Colloquium 180}, ed. by K. J. Johnston, Washington, D.C., 
254--258, 2000a.

\bibitem[{\it Petrov}(2000b)]{r:petinst} 
   Petrov,~L., Instrumental errors of geodetic VLBI,
{\it Proc. of the in International VLBI service for geodesy and astrometry 
2000 general meeting}, ed. by N.~Vandenberg and K.~Baver, Greenbelt, 230--235, 
2000b.

\bibitem[{\it Petrov and Ma}(2002)]{r:harpos} 
   Petrov,~L. and C.~Ma, Study of harmonic site position variations determined 
by VLBI, \jgr {\it 108}(B4), 2190, doi: 10.1029/2002JB001801, 2003.

\bibitem[{\it Ponte and Ray}(2002)]{r:ponteray} 
   Ponte,~R.M. and Ray,~R.D., Atmospheric pressure correction in geodesy and
oceanography: A strategy for handling air tides, \grl, {\it 29}(24), 2153,
doi:10.1029/2002GL016340, 2002.

\bibitem[{\it Rabel and Zschau}(1985)]{r:rab}
   Rabbel,~W. and J. Zschau, Static deformation and gravity changes at the
earth's surface due to atmosphere loading, {\it J. Geophys.}, {\it 56}, 
81--99, 1985.

\bibitem[{\it Rabel and Schuh}(1986)]{r:rabshu}
  Rabbel,~W. and H.~Schuh, The influence of atmosphere loading on 
VLBI-experiments, {\it J. Geophys.}, {\it 59}, 164--170, 1986.

\bibitem[{\it Ray}(1999)]{r:got99} 
   Ray,~R.D., A global ocean tide model from TOPEX/POSEIDON Altimetry: 
GOT99.2, {\it NASA/TM-1999-209478}, Greenbelt, 58~p., 1999.



\bibitem[{\it Sovers et al.}(1998)]{r:masterfit} 
   Sovers,~O.J, J.L. Fanselow, and C.S. Jacobs, Astrometry and geodesy with
radio interferometry: experiments, models, results, {\it Reviews of Modern 
Physics}, {\it 70}(4), 1393--1454, 1998.

\bibitem[{\it Takahashi et al.}(2000)]{r:ksp} 
   Takahashi,~F., T.~Kondo, Y.~Takahashi, and Y.~Koyama, Very Long Baseline
Interferometer, Ohmsha, Ltd, Tokyo, 243~p., 2000.

\bibitem[{\it Tierney et al.}(2000)]{r:tierney}
   Tierney,~C., J.~Wahr, F.~Bryan and V.~Zlotnicki, Short-period oceanic 
circulation: implications for satellite altimetry, \grl {\it 27}, 1255--1258, 
2000.

\bibitem[{\it Trubytsyn and Makalkin}(1976)]{r:truba} 
   Trubytsyn,~A.P. and A.V.~Makalkin, Deformations of the Earth's crust due to
atmospheric cyclones, Izvestia of Academy of Sciences USSR, Phys. Solid Earth,
{\it 12}, 343--344, 1976.

\bibitem[{\it van Dam and Herring}(1994a)]{r:damvlbi} 
  van Dam,~T.M. and T.A.~Herring, Detection of atmospheric pressure loading 
using Very Long Baseline Interferometry measurements, \jgr {\it 99},
4505--4518, 1994a.

\bibitem[{\it van Dam et al.}(1994b)]{r:gpsatmloa} 
   van Dam,~T.M., G.~Blewitt, and M.~Heflin, Detection of atmospheric 
pressure loading using the Global Positioning System, \jgr {\it 99}, 
23,939--23,950, 1994b.

\bibitem[{\it van Dam and Wahr}(1987)]{r:dampre} 
   van Dam,~T.M. and J.~Wahr, Displacements of the Earth's surface due to 
atmospheric loading: Effects on gravity and baseline measurements, \jgr
{\it 92}, 1281--1286, 1987.


\bibitem[{\it van den Dool et al.}(1997)]{r:dool} 
   van~den~Dool,~H.M., S.~Saha, J.~Schemm, J.~Huang,
A temporal interpolation method to obtain hourly atmospheric surface pressure
tides in Reanalysis 1979-1995, \jgr {\it 102}(D18), 22,013--22,2024, 1997.

\bibitem[{\it Vandenberg}(1999)]{r:ivs} 
   International VLBI Service for Geodesy and Astrometry, 1999 annual report,
ed. N.~Vandenberg, {\it NASA/TP-1999-209243}, Greenbelt, USA, 308~p., 1999.

\bibitem[{\it Velicogna et al.}(2001)]{r:velicogna}
   Velicogna, I., J. Wahr and H. van den Dool, Can surface pressure be used
to removed atmospheric contributions from GRACE data with sufficient accuracy
to recover hydrologic signals?, \jgr {\it 106}(B8), 16,415--16,434, 2001. 

\bibitem[{\it Wunsch and Stammer}(1997)]{r:wunsch}
   Wunsch,~C. and D.~Stammer, Atmospheric loading and the "inverted barometer" 
effect, {\it Rev. Geophys.}, {\it 35}, 117--135, 1997.

\end{thebibliography}
\end{document}